\documentclass{ws-mpla}
\usepackage[super]{cite}
\usepackage{graphicx}
\usepackage{hyperref}
\usepackage{graphicx}
\usepackage{graphbox}

\def\pt{$p_{\text{T}}$}

\def\met{$E_{\text{T}}^{\text{miss}}$}
\def\mt{$m_{\text{T}}$}
\def\meff{$m_{\text{eff}}$}

\def\ttbar{$t\bar{t}$}

\def\GeV{$\textnormal{GeV}$}
\def\ifb{$\textnormal{fb}^{-1}$}
\newcommand{\mttwo}{\ensuremath{m_\text{T2}}}

\def\TeV{\ifmmode {\mathrm{\ Te\kern -0.1em V}}\else
                   \textrm{Te\kern -0.1em V}\fi}%
\def\GeV{\ifmmode {\mathrm{\ Ge\kern -0.1em V}}\else
                   \textrm{Ge\kern -0.1em V}\fi}%
\def\MeV{\ifmmode {\mathrm{\ Me\kern -0.1em V}}\else
                   \textrm{Me\kern -0.1em V}\fi}%                  

\newcommand{\ninoone}{\ensuremath{\tilde{\chi}_{1}^{0}}}
\newcommand{\ninotwo}{\ensuremath{\tilde{\chi}_{2}^{0}}}
\newcommand{\chinoone}{\ensuremath{\tilde{\chi}_{1}^{\pm}}}

\begin{document}

\markboth{Jeanette Miriam Lorenz}
{Supersymmetry and the collider Dark Matter picture}

%%%%%%%%%%%%%%%%%%%%% Publisher's Area please ignore %%%%%%%%%%%%%%
\catchline{}{}{}{}{}
%%%%%%%%%%%%%%%%%%%%%%%%%%%%%%%%%%%%%%%%%%%%%%%%%%%%%%%%%%%%%%%%%%%

\title{Supersymmetry and the collider Dark Matter picture}

\author{\footnotesize Jeanette Miriam Lorenz}

\address{Fakult\"at f\"ur Physik, LMU Munich, Am Coulombwall 1\\
D-85748 Garching\\
Germany\\
Jeanette.Lorenz@physik.uni-muenchen.de}

\maketitle

\pub{Received (19.08.2019)}{Revised (26.08.2019)}

\begin{abstract}

One of the key questions in particle physics and astrophysics is the nature of dark matter, which existence has been confirmed in many astrophysical and cosmological observations. Besides direct and indirect detection experiments, collider searches for dark matter offer the unique possibility to not only detect dark matter particles but in case of discovery to also study their properties by making statements about the potential underlying theory. The search program for dark matter at the ATLAS and CMS experiments at the Large Hadron Collider is comprehensive, and includes both supersymmetric dark matter candidates and other alternatives. This review presents the latest status in these searches, with special focus on supersymmetric dark matter particles.

\end{abstract}

\section{Introduction}

The Standard Model (SM) of particle physics describes matter and its interactions -- except gravity -- to a remarkable precision. Still, good reasons exist to believe that the SM is an incomplete theory only valid at low energies. For instance, it does not offer a candidate for Dark Matter (DM): the non-luminous and non-absorbing matter in universe.

The existence of Dark Matter is inferred from different cosmological and astrophysical observations\cite{PDGDMreview,Bertone}, e.g. from the motion or rotational curves of luminous objects such as stars, gas clouds, globular clusters and galaxies. One of the prominent recent examples is the transit of the Bullet Cluster through another cluster\cite{Clowe:2006eq} where gravitational lensing showed that the total mass of the clusters moved faster than the luminous hot gas decelerated. Taking the different observations together, in particular also including the measurements of the anisotropy in the cosmic microwave background\cite{Akrami:2018vks,Aghanim:2018eyx}, the density of cold, non-baryonic dark matter is measured as $\Omega_{\mathrm{nbm}}h^{2} = 0.1186 \pm 0.0020$, where $h$ is the Hubble constant\cite{PDGDMreview_cosmological}.

Several conditions need to be full-filled by candidates for non-baryonic DM\cite{PDGDMreview}: it needs to be stable on cosmological scales, may not emit or absorb electromagnetic radiation (else it would not be non-luminous) and needs to result in the correct relic density (the density at the time of the freeze-out). The last point is indeed a criterion not trivially achieved by the different DM candidates proposed by Beyond-the-Standard-Model (BSM) theories.

Widely studied candidates for DM are Weakly Interacting Massive Particles (WIMPs). WIMPS appear in multiple BSM theories such as in \textit{Little Higgs} models or \textit{Technicolor}, but one of the best candidates is the lightest supersymmetric particle (LSP) appearing in \textit{supersymmetric theories} (SUSY) (see discussion in Ref.~\refcite{PDGDMreview}), if it is stable through strictly enforcing R-parity conservation\cite{DMSUSYreview}, forbidding its decay to SM particles.
Other DM candidates are e.g. primordial back holes, axions and sterile neutrinos.

\begin{figure}
 \centerline{
  \includegraphics[width=0.35\textwidth]{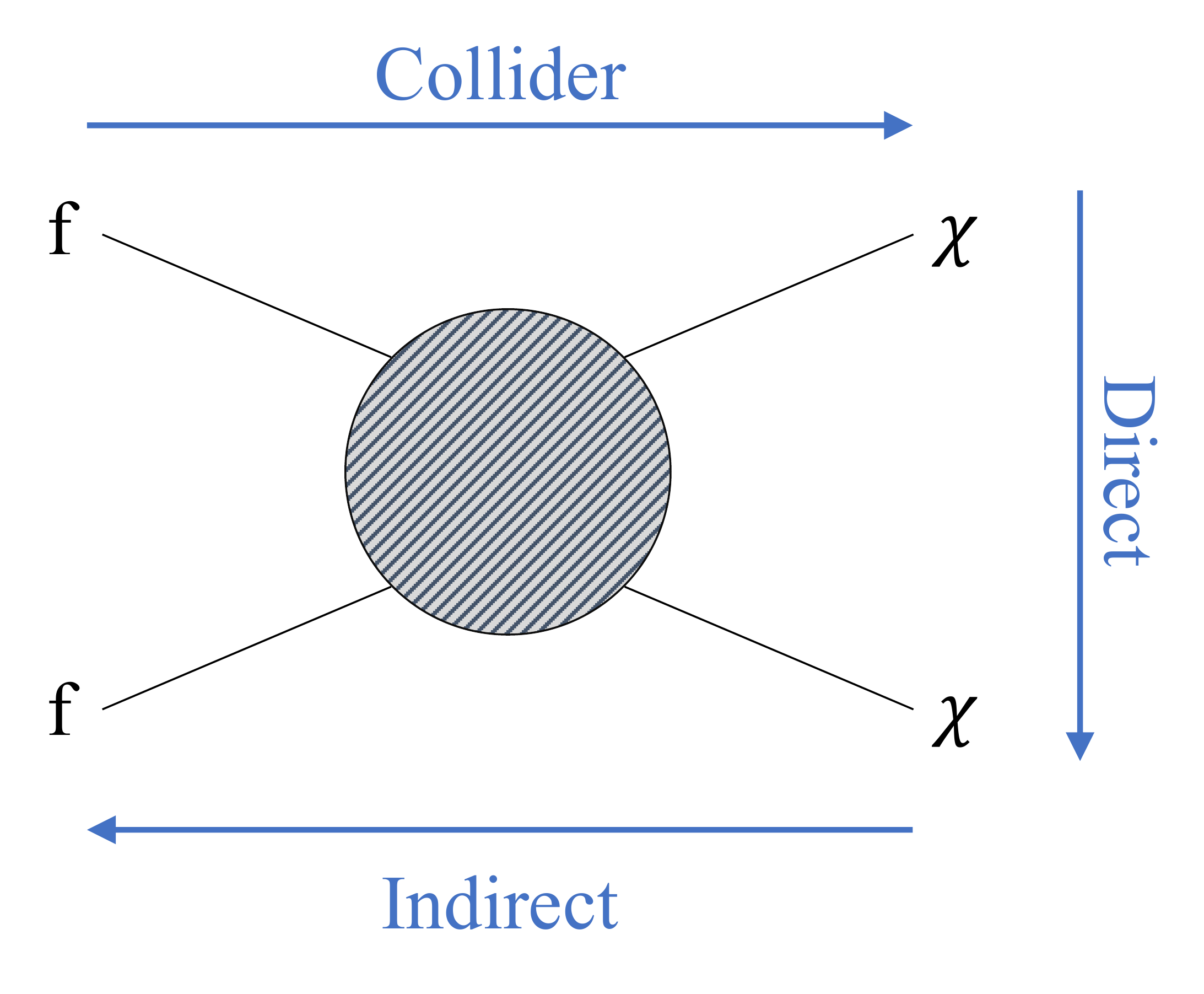}
 }\vspace*{8pt}
\caption{\label{DM_experiments}Interplay\cite{ATLAS_DMsummary} of different dark matter experiments: Direct detection experiments search for collision of galactic DM with material in underground detectors. Indirect detection experiments look for annihilation of galactic DM particles. Collider searches aim to produce DM particles.}
\end{figure}

Searches for DM, and in particular WIMPs, are performed by different classes of experiments as shown in Fig.~\ref{DM_experiments}. Direct detection (DD) experiments seek to discover collisions of galactic DM with material in underground detectors, while indirect detection experiments (ID) look for products of annihilating DM~\cite{Bertone}. Particle colliders such as the Large Hadron Collider (LHC)\cite{lhc}, with the two general-purpose detectors ATLAS\cite{atlas1,atlas2,atlas3} and CMS\cite{cms}, aim to produce DM particles in collisions of SM particles. All these experiments have put constraints on WIMPs over the last years\cite{PDGDMreview}. Nonetheless, searches for supersymmetric WIMPs in particular still have a lot of room for discovery, as up to now the most stringent constraints on SUSY particles are on strongly interacting SUSY particles.

Different models for the production of DM particles are possible at colliders, as detailed in Sec.~\ref{DMmodels}. In case of SUSY, the DM particles often appear in decays of other, heavier, SUSY particles (if not considering the direct production of LSPs). Both ATLAS and CMS pursue a comprehensive search program for SUSY particles, as summarized with a few examples in Sec.~\ref{SUSYsearches}. Searches for SUSY are the main focus of this review.
Other DM models used at colliders (which are referred to as \textit{mediator simplified models} in this review to distinguish them from SUSY simplified models) make as few assumptions as possible on the underlying theory, as detailed in Sec.~\ref{DMmodels}. In this case, the direct production of a pair of DM particles is assumed, appearing in the decay of a mediator particle connecting SM and DM particles. A few example searches are illustrated in Sec.~\ref{monoX}. A third class of models considers the direct decay of Higgs or $Z$ bosons into DM particles (invisible Higgs- and $Z$ boson decays). An example for searches for invisible Higgs decays is given in Sec.~\ref{SM}. A more extensive recent review on the non-SUSY DM searches at the LHC is given in Ref.~\refcite{doglioni}. The searches,  with different targets and interpretations, are complementary and provide a wide coverage of possible DM scenarios at colliders. A comparison of collider searches with DD experiments is given in Secs.~\ref{SUSYDMDD} and \ref{DMsummary}.

The LHC, ATLAS and CMS had two successful data-taking periods. In the first period (Run-1) from 2010-2012, $pp$-collisions at a center-of-mass energy of 7 and 8 \TeV\ resulted in a total statistics of about 5 \ifb\ at 7 \TeV\ and 20 \ifb\ at 8 \TeV. After a technical stop with upgrades to the collider and detectors, $pp$-collisions at a center-of-mass energy of 13 \TeV\ resumed in 2015 - 2018. Due to the only recent end of this second data-taking period (Run-2) many analyses at ATLAS and CMS do not use the full dataset yet, but showed results with the partial 2015-2016 dataset amounting to about 36 \ifb\ or the partial 2015-2017 dataset with about 80 \ifb. Only recently, the first results using the full dataset of $\sim$ 140 \ifb\ were presented by a small number of analyses. The dataset of Run-2 is significantly larger and at higher energy than the dataset in Run-1 and thus allows -- as shown in the further course of this review -- to perform certain searches for electroweakly produced SUSY particles for the first time at LHC experiments.

\section{Models in DM searches}
\label{DMmodels}

Different models, depending on the underlying theory, are used in the design and interpretation of searches for DM particles at colliders.

\subsection{Supersymmetry}
\label{SUSY}

Supersymmetry\cite{Martin_primer} introduces an additional symmetry between fermions and bosons and extends space-time symmetry. Of the different supersymmetric theories possible, only the low-energy realization of $N=1$ SUSY might be accessible to present colliders. The most common considered model is the Minimal Supersymmetric Standard Model (MSSM)\cite{Fayet:1976et,Fayet:1977yc} which is the minimal possible extension of the SM by SUSY particles. Every SM particle has SUSY superpartner(s) (sparticles). These sparticles have the same properties as their SM partners, except of the spin which differs by $\pm 1/2$ and mass (as SUSY needs to be softly broken). Superpartners to leptons are sleptons ($\tilde{l}$), to quarks squarks ($\tilde{q}$), to the gluon gluinos ($\tilde{g}$). As a superpartner is required for both left- and right-handed quarks and leptons, each of them has two superpartners $\tilde{q}_{\mathrm{L}}$ and $\tilde{q}_{\mathrm{R}}$ or $\tilde{l}_{\mathrm{L}}$ and $\tilde{l}_{\mathrm{R}}$, respectively. Third-generation superpartners (of the top, bottom, and tau) mix, forming a lighter and a heavier state. E.g. two stops are obtained with the lighter being $\tilde{t}_{1}$ and the heavier $\tilde{t}_{2}$. 

In SUSY two complex Higgs doublets are required to give mass to down-type quarks and leptons or up-type quarks, respectively. After spontaneous symmetry breaking, five Higgs bosons are obtained: two CP-even neutral $h$ and $H$, where the lighter $h$ is often assumed to correspond to the SM Higgs boson, while the $H$ is heavier; a CP-odd $A$ boson; and two charged Higgs bosons $H^{\pm}$. At tree level the Higgs sector depends, beside the coupling constants to matter, on $\tan \beta = \frac{v_{1}}{v_{2}}$, with $v_{1}$ and $v_{2}$ being the vacuum expectation values of the two Higgs doublets, and on one Higgs boson mass, e.g. $m_{A}$.
Superpartners to the Higgs bosons are higgsinos: $\tilde{H}_{\mathrm{u}}^{0}$, $\tilde{H}_{\mathrm{d}}^{0}$, $\tilde{H}_{\mathrm{d}}^{-}$ and $\tilde{H}_{\mathrm{u}}^{+}$. They mix with the superpartners of the $W$- and $B$-fields, the winos ($\tilde{W}^{0}$, $\tilde{W}^{\pm}$) and bino ($\tilde{B}$) to the physical mass eigenstates charginos ($\tilde{\chi}_{1,2}^{\pm}$) and neutralinos ($\tilde{\chi}_{1,2,3,4}^{0}$). 

Supersymmetric theories in general may contain interactions which allow proton decay. One possibility to prevent this unobserved process is to introduce an additional conserved, multiplicative, quantum number, the $R$-parity\cite{Farrar:1978xj} $R = (-1)^{3(B-L)+2s}$, with $B$ being the baryon number, $L$ the lepton number and $s$ the spin. The $R$-parity is $R = -1$ for SUSY particles and $R = +1$ for SM particles. Introducing $R$-parity has important consequences for the phenomenology, as SUSY particles can then only be produced pair-wise and decay until the lightest supersymmetric particle is obtained. $R$-parity enforces that this particle is stable. Cosmological constraints require the LSP to be electrically and color neutral\cite{Ellis:1983ew}. Due to its stability it is an excellent DM candidate~\cite{Goldberg:1983nd,Ellis:1983ew}. All SUSY searches presented in this review assume $R$-parity conservation.

Different candidates for the LSP, such as e.g. the sneutrino (the supersymmetric partner of the neutrino) or the lightest neutralino, exist in SUSY theories. The sneutrino is however mostly ruled out in the MSSM, as the annihilation rate would be too high, and the regions in which the correct relic density is obtained are already excluded\cite{sneutrino1,sneutrino2}. The neutralino is instead an attractive candidate. The requirement to achieve the correct relic density puts constraints on the type of the neutralino\cite{PDGDMreview}. It could be e.g. mostly a bino if both the LSP mass and a slepton mass are below $\sim$ 150 \GeV\cite{PDGDMreview}, or the masses are close so that co-annihilation between LSP and a slepton occurs, or $2 m(\mathrm{LSP}) \sim m(h)$. The LSP could also be mostly a higgsino or wino. In case of the LSP being purely wino or higgsino, calculations of the relic density predict masses of about $1 \TeV$ for higgsinos or $3 \TeV$ for winos\cite{DMwino1,DMwino2}. Other SUSY particles may also compose DM, but are extremely challenging to detect: the only gravitationally interacting gravitino (the superpartner of the graviton), not being in thermal equilibrium in the early Universe\cite{Bertone}, and the axino (the superpartner of the axion).

SUSY is also able to address other short-comings of the SM\cite{Martin_primer} such as the so-called hierarchy problem, and possibly achieves gauge unification at high energies. The hierarchy problem arises as the mass of the SM Higgs boson -- discovered in 2012\cite{atlas_higgs,cms_higgs} -- turns out to be at the electroweak scale at 125 \GeV, while radiative corrections to its mass depend on the cut-off scale until which the theory is valid. This cut-off scale might in the worst case be of the size of the Planck scale at $10^{19}$ \GeV, raising the mass of the Higgs boson to this scale as well. If these radiative corrections are not compensated through means of a BSM theory, so-called fine-tuning of the parameters has to occur to keep the Higgs boson at its measured mass (this is also called the \textit{naturalness} problem\cite{Barbieri:1987fn,deCarlos:1993yy,Feng:2013pwa}).

\subsubsection{SUSY models}

In performing a search for sparticles, it is not feasible to design the search assuming the full MSSM, or interpret in the MSSM directly, due to the large number of free parameters (124 parameters). Various simplifications of the MSSM have been proposed and are used in searches. The phenomenological MSSM (pMSSM)\cite{Djouadi:1998di,Berger:2008cq} reduces the free parameters to 19 parameters. These parameters include the masses of the gauginos (superpartners of the gauge bosons): $M_{1}$ for the Bino, $M_{2}$ for the Wino and $M_{3}$ for the gluino; the parameters determining the Higgs sector with $\tan{\beta}$, $m_{\mathrm{A}}$ and the higgsino mass parameter $\mu$. A further five parameters describe the sfermion masses of the first two generations, which are taken as degenerate, and other five parameters the sfermion masses of the third generation. The remaining parameters are connected to the coupling of the third-generation sfermions to the Higgs sector.
A summary of searches interpreted in the pMSSM was presented by ATLAS for the Run-1 searches\cite{ATLAS_pMSSM}.

Most searches design their analysis using simplified models\cite{Alwall:2008ve,Alwall:2008ag,Alves:2011wf}
in which the production of specific sparticles is assumed with a clearly defined decay to lighter sparticles and eventually the LSP. All other sparticles not appearing explicitly in this decay are assumed to be heavy enough so that they decouple. A few simplified models are shown in Fig.~\ref{SUSY:simplified}.

\begin{figure}
 \centerline{
  \includegraphics[width=0.24\textwidth]{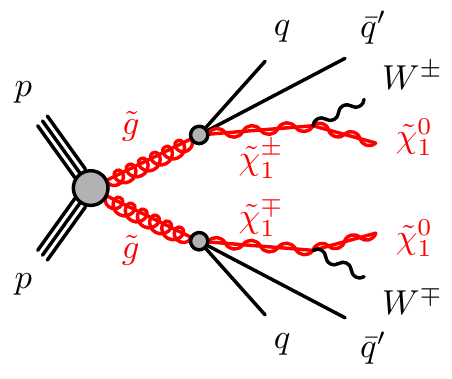}
  \includegraphics[width=0.24\textwidth]{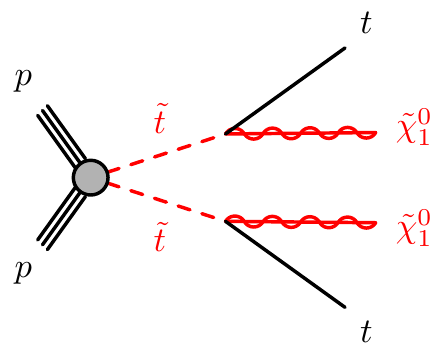}
  \includegraphics[width=0.24\textwidth]{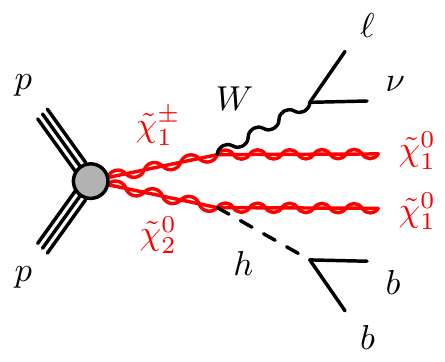}  
 }\vspace*{8pt}
\caption{\label{SUSY:simplified} Different example simplified models\cite{0lepton,stop_1L,ATLAS_Wh} used in SUSY searches: A model with gluino pair production and decays to LSPs via charginos (left), with stop pair production and direct decays to tops and LSPs (middle) and chargino/neutralino pair production with decays to Higgs and $W$ bosons and LSPs (right). A big circle in contrast to a normal vertex indicates that an additional interaction including a virtual particle occurred.}
\end{figure}

While being practical for the design and interpretation of a given search, the simplified models do not present a complete supersymmetric theory. Search results based on simplified models can however be interpreted in complete SUSY theories, as e.g. shown in Ref.~\refcite{gambit_paper}.

\subsection{Other models}

\subsubsection{Mediator simplified models}

Without specifying the underlying theory concretely, mediator simplified models can be used in the search for DM particles. These special simplified models assume the direct production of DM particles in the decay of a BSM mediator particle connecting SM and DM particles. A few examples are shown in Fig.~\ref{monoX_diagrams}.
Different simplified models have been suggested by the LHC DM WG\cite{Albert:2017onk,Boveia:2016mrp,Abe:2018bpo,doglioni}. As in the case of SUSY simplified models, it is assumed that any other new physics is at much higher energies than the energy scale accessible to the experiment, so that only the DM particles $\chi$ and the BSM mediator particle appear in addition to SM particles.

An assumption often made for the mediator particle is a spin-1 vector- or axial-vector $Z'_{V/A}$ boson, which results from an extension of the SM by an additional U(1) gauge symmetry (vector- or axial-vector simplified model) as in Fig.~\ref{monoX_diagrams}.
Such a $Z'_{V/A}$ appears in many BSM extensions. As the $Z'_{V/A}$ couples to both SM and DM particles, decays into both particle types are possible, resulting in two different classes of searches for the mediator: either searching for the decays of the $Z'_{V/A}$ into DM particles or into SM particles. The latter are so-called \textit{mediator} searches where often decays of the $Z'_{V/A}$ into two quarks are considered, resulting in signatures with two possibly close-by jets, depending on the boost of the $Z'_{V/A}$ (see e.g. Ref.~\refcite{ATLAS_DMsummary}). Mediator searches are not discussed in this review. The former class of searches has historically been referred to as mono-X searches, as typically searching for a SM particle (or few SM particles) in association with missing energy from invisible decay products (since the DM particle are invisible to the detector, as only weakly or gravitationally interacting). 

Different parameter values need to be chosen to define the vector- or axial-vector simplified model: the masses of the mediator particle ($m_{Z'_{V,A}}$) and DM particle ($m_{\chi}$), the common couplings of the mediator particle to any quarks ($g_{q}$), to leptons ($g_{L}$) and to DM particles ($g_{X}$).

Other simplified models are also considered by collider searches, such as a mediator which is a (pseudo-)scalar particle or a spin-2 particle, possibly leading to different signatures. A recent summary of models including constraints is given in Ref.~\refcite{ATLAS_DMsummary}.

\begin{figure}
 \centerline{
  \includegraphics[width=0.24\textwidth]{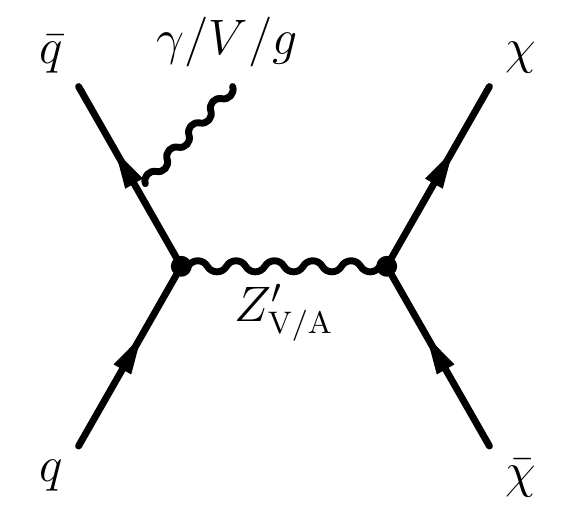}
  \includegraphics[width=0.24\textwidth]{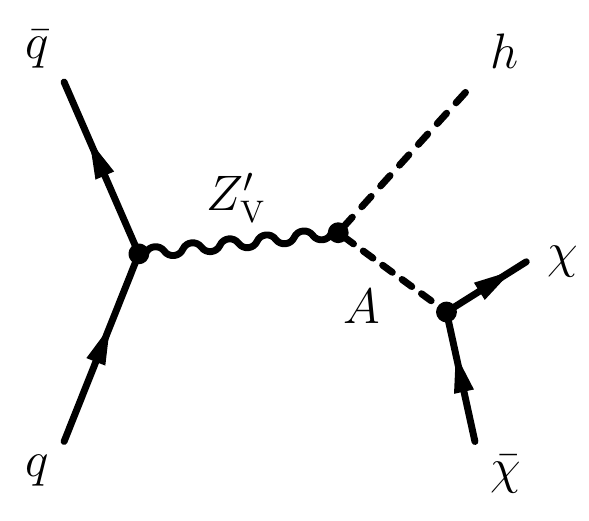}
  \includegraphics[width=0.24\textwidth]{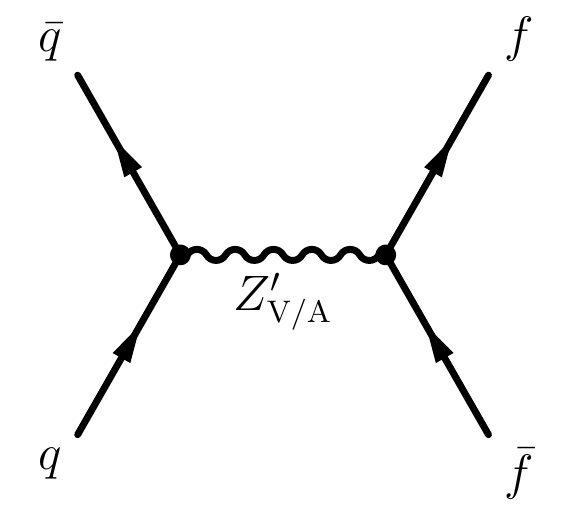}
 }\vspace*{8pt}
\caption{\label{monoX_diagrams}Different simplified models\cite{ATLAS_DMsummary} for the production of fermion-like dark matter particles $\chi$ via a BSM mediator particle. Left: with initial-state radiation of a SM particle; middle: with emission of a SM particle from the BSM mediator particle. An alternative in searches for DM particles are searches for decays of the mediator particle to SM particles (right).}
\end{figure}

\subsubsection{Higgs/Z-portal dark matter models}

In Higgs and $Z$-portal DM models\cite{Shrock:1982kd}, the Higgs or $Z$ boson acts as mediator between the SM and the DM particles. The DM particles are thus the only addition on top of the SM content. The Higgs or $Z$ bosons might directly decay to a pair of DM particles, if their mass is below half the mass of the Higgs or $Z$ boson. This results in invisible Higgs or $Z$ decays. LEP and DD experiments put strong constraints on $Z$-portal models\cite{Escudero:2016gzx}. In case of heavier DM particles, invisible decays would be forbidden, but might indirectly influence the decay rates of Higgs and $Z$ bosons to other particles. 

\subsubsection{Effective field theory}

Effective-field-theory models (EFTs)\cite{Goodman:2010ku,Fox:2011pm} neglect the precise type of coupling or mediator type between SM and DM particles and just assumes a four-point contact interaction. EFTs are valid if the mass of the mediator is much higher than the collision energy. The advantage of these models is that any higher-energy details of the interaction are not of interest for searches. 
If however the mass of the mediator between SM and DM particles is smaller or in the size of the energy scale of the collision, EFTs are not valid. The application of EFTs in DM searches at colliders is discussed in Refs.~\refcite{Busoni:2013lha,Racco:2015dxa}.

\section{Concepts of collider DM searches}
\label{concepts}

Analysis strategies in searches for DM particles at colliders are mostly very similar.
In the following, the typical concept and analysis strategy of such a search is illustrated by the example of a SUSY search for the pair production of gluinos.

The decay of a gluino proceeds via a cascade decay to the LSP with one option depicted in Fig.~\ref{SUSY:simplified}, left. In all cases $\tilde{g} \rightarrow \tilde{q}q^{'}$ and then e.g. $\tilde{q} \rightarrow \tilde{\chi}^{\pm}_{1} q$. For  $m(\tilde{q}) > m(\tilde{g})$ the decay of the gluino appears as 3-body decay: $\tilde{g} \rightarrow \tilde{\chi}^{\pm}_{1} qq^{'}$. The $\tilde{\chi}^{\pm}_{1}$ decays further as $\tilde{\chi}^{\pm}_{1} \rightarrow W^{\pm} \tilde{\chi}^{0}_{1}$. The LSP $\tilde{\chi}^{0}_{1}$ escapes the detector as being stable and only weakly-interacting. It thus leads to missing momentum in the transverse plane of the $pp$-collision (perpendicular to the beam directions).

This missing transverse momentum (\met\ or $p^{\mathrm{miss}}_{\mathrm{T}}$) is an important signature common to many searches for DM particles at the LHC. Also SM processes may lead to \met\ if a neutrino is produced, but typically \met\ is considerably higher in BSM signals. Besides this \textit{true} \met,  mis-measurements of e.g. jet energies in the calorimeter or other in-time collisions (pile-up) may lead to misidentified -- \textit{fake} -- \met. It is thus of importance to guarantee a good understanding and reconstruction of \met\ -- in particular regarding pile-up -- to ensure a suppression of fake \met\ and excellent reconstruction of true \met. Due to differences in their detector and reconstruction methods CMS and ATLAS reconstruct the \met\ differently\cite{cms_met,met_paper}. In case of ATLAS, \met\ is defined by\cite{met_paper}:

\begin{eqnarray}
 E_{x(y)}^{\mathrm{miss}} &=& - \sum_{i \in \{ \mathrm{hard~objects} \}} p_{x(y),i} - \sum_{j \in \{ \mathrm{soft~signals} \}} p_{x(y),j} \label{met}\\
 E_{\mathrm{T}}^{\mathrm{miss}} &=& | \mathbf{p}_{\mathrm{T}}^{\mathrm{miss}} | = \sqrt{ (E_{x}^{\mathrm{miss}})^2 + (E_{y}^{\mathrm{miss}})^2}
\end{eqnarray}

\noindent where the first term in Eq.~\ref{met} is the \textit{hard term} and consists of the measured transverse momenta of all reconstructed particles or objects such as electrons, photons, taus, muons and jets. The second term collects the transverse momenta of all tracks not contributing to the reconstructed objects in the first term and is consequently very sensitive to pile-up effects. Various algorithms for suppression of pile-up are used\cite{met_paper}. CMS constructs the \met\ from particle-flow objects\cite{cms_met}. Both CMS and ATLAS see an excellent performance in the \met\ reconstruction as illustrated in Fig.~\ref{met_performance}.

\begin{figure}
 \centerline{
  \includegraphics[align=t,width=0.48\textwidth]{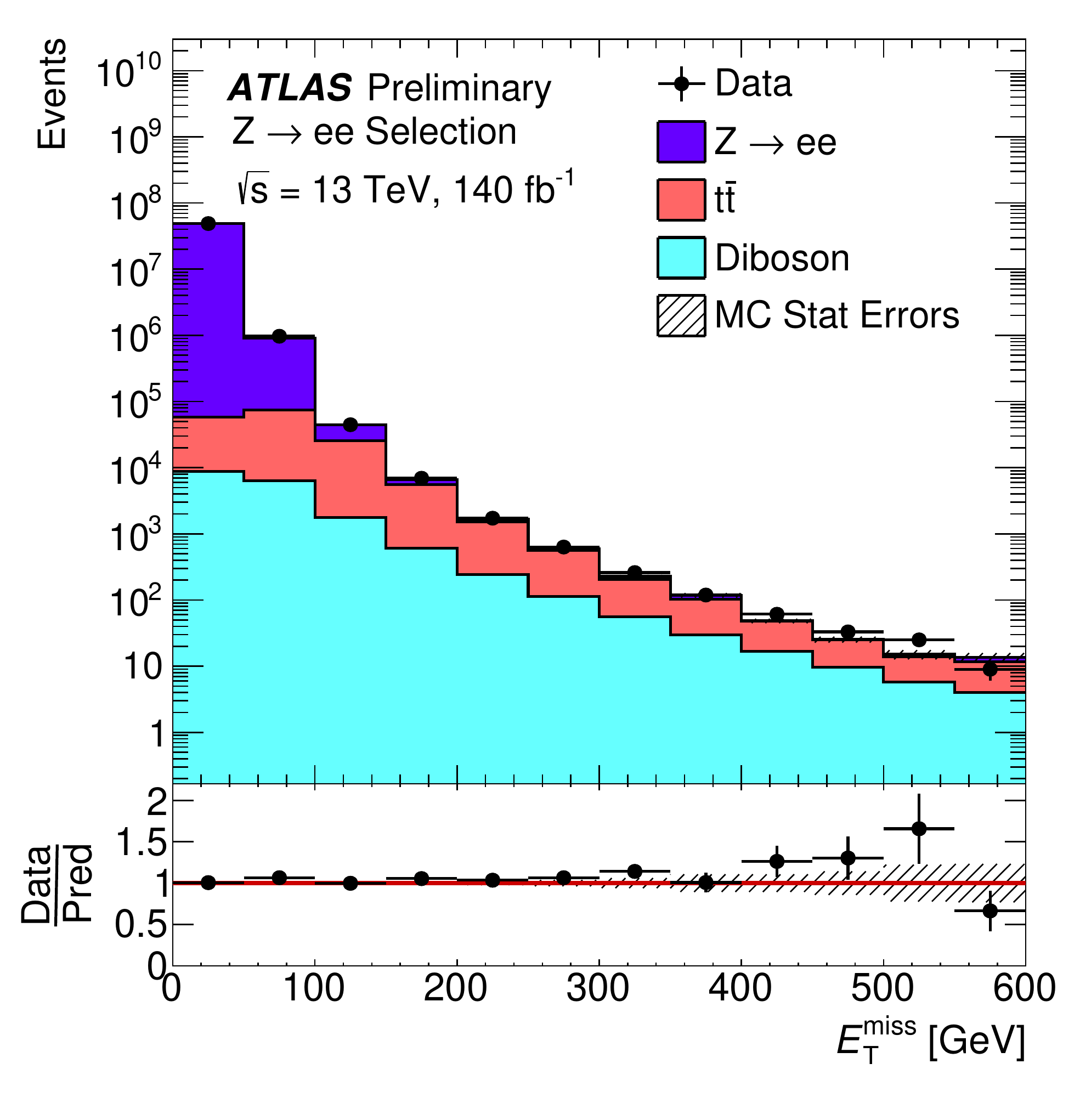}
  \includegraphics[align=t,width=0.495\textwidth]{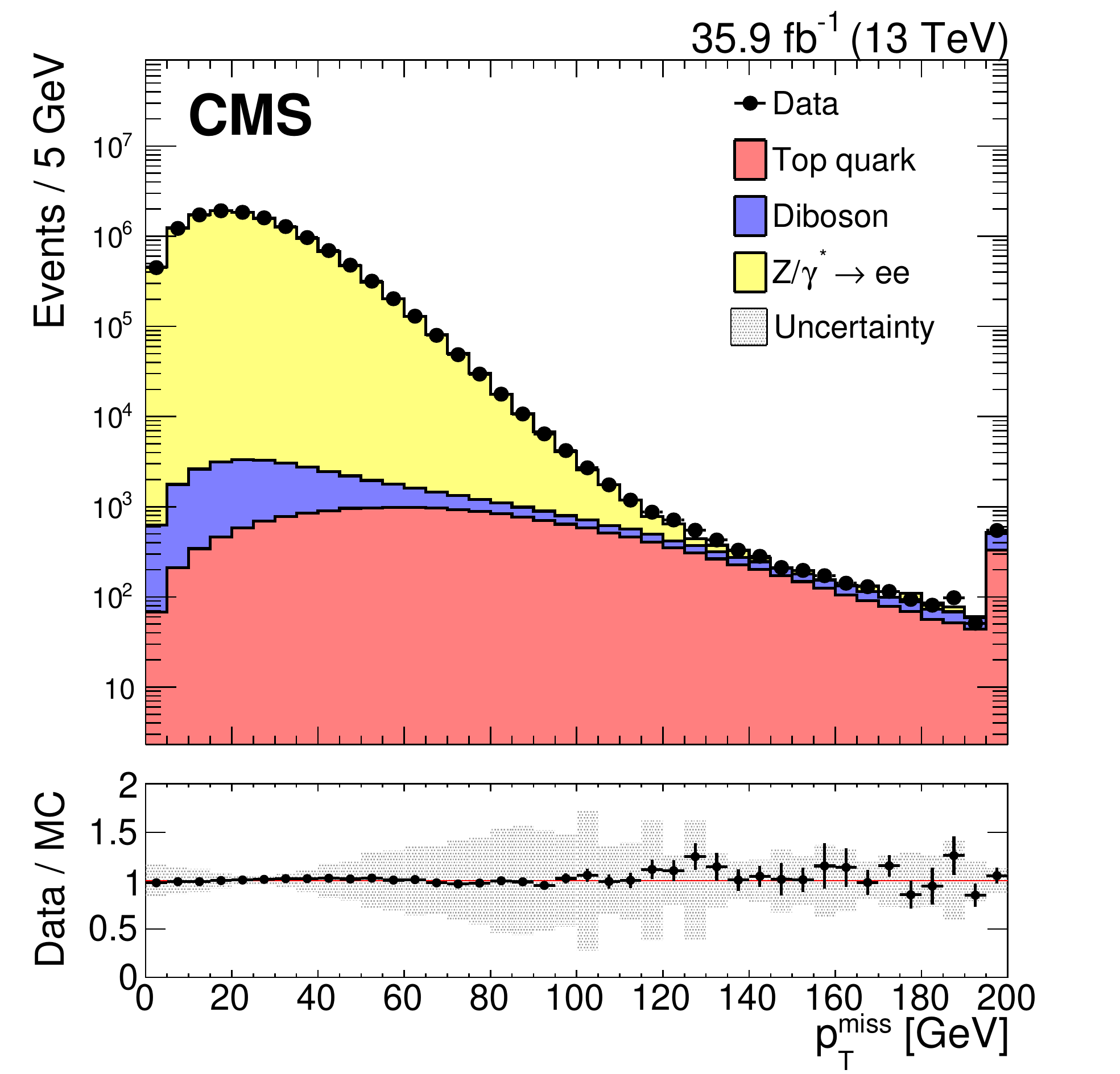}
  }
  \vspace*{8pt}
\caption{\label{met_performance}In $Z \rightarrow e^{+}e^{-}$ processes no true \met\ is present, but fake \met\ can occur. Distributions of \met\ are shown for ATLAS\cite{met_PUB} (left) and for CMS\cite{cms_met} (right) in such a selection. Data is compared to MC simulation of the most important processes contributing such as $Z \rightarrow e^{+}e^{-}$, \ttbar\ and diboson. The good agreement shows the excellent performance obtained in the \met\ reconstruction.}
\end{figure}

In addition to \met, the above described decay of gluino pairs results in two $W$ bosons and in multiple jets from the emitted quarks. The signatures searched for thus consist of multiple jets, \met\ and possibly leptons from further decays of the $W$ bosons. A typical analysis will use various observables constructed from the properties of the decay products to enhance the signal while suppressing SM backgrounds. Two observables frequently used in SUSY searches are shown in Fig.~\ref{SUSY:kinematics}. 

\begin{figure}
 \centerline{
  \includegraphics[width=0.44\textwidth]{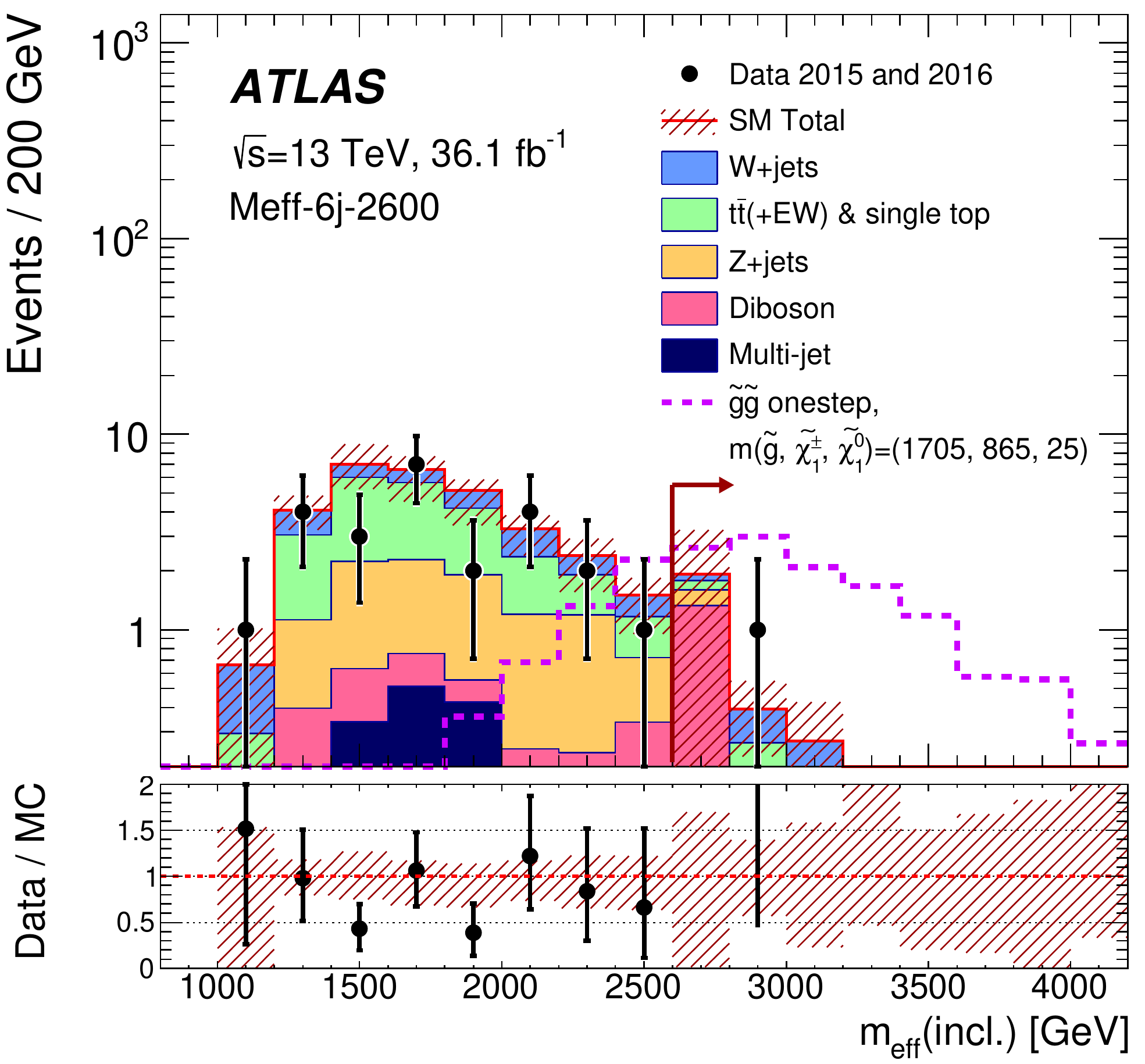}
  \includegraphics[width=0.54\textwidth]{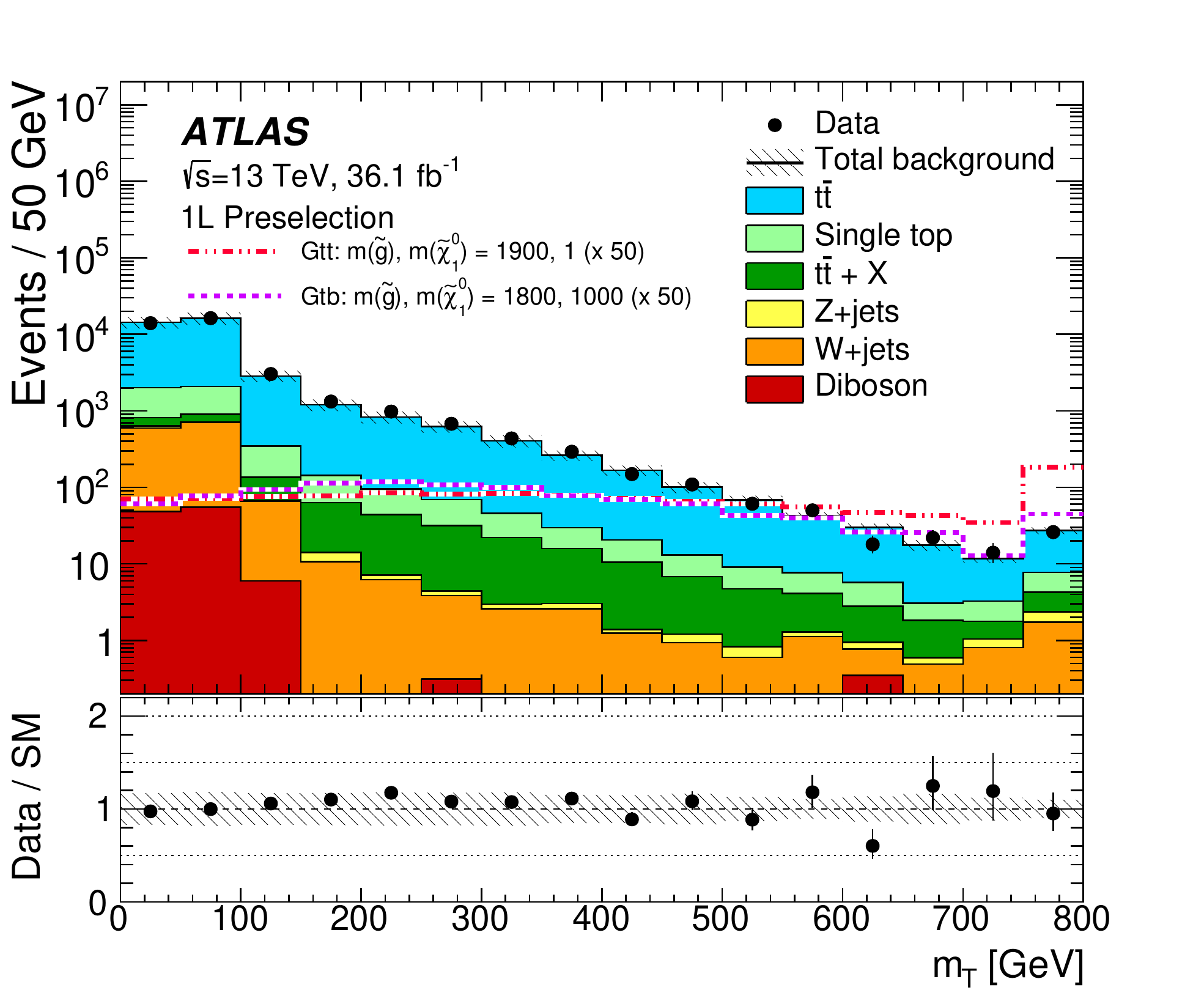}
  }
  \vspace*{8pt}
\caption{\label{SUSY:kinematics}Examples for observables useful in searches for supersymmetric particles. The \meff\ is shown for an analysis addressing fully hadronic signatures\cite{0lepton} (left). The \mt\ is presented for an analysis selecting events with at least three b-tagged jets and an electron or a muon\cite{multib}.}
\end{figure}

The effective mass (\meff) is correlated with the mass scale of the primary SUSY particle\cite{Hinchliffe:1996iu,Tovey:2000wk,Cabrera:2012cj}:

\begin{equation}
 m_{\mathrm{eff}} = \sum_{\mathrm{jets}} p_{\mathrm{T}} + \sum_{\mathrm{leptons}} p_{\mathrm{T}} + E_{\mathrm{T}}^{\mathrm{miss}}
\end{equation}

\noindent The transverse mass (\mt) was originally proposed in searches for $W$ bosons and shows a sharp cut-off at the mass of the $W$\cite{Wmt1,Wmt2}. 
In searches for SUSY with a lepton in the signature, the signal often shows higher values than the background due to higher \met\ values in the signal (see Ref.~\refcite{Barr:2009jv} for a summary on different definitions of transverse mass quantities):%potentially more references

\begin{equation}
 m_{\mathrm{T}} = \sqrt{2 p_{\mathrm{T}}^{\mathrm{lepton}} E_{\mathrm{T}}^{\mathrm{miss}} (1 - \cos{[\Delta \Phi(\mathbf{p}_{\mathrm{T}}^{\mathrm{miss}},\mathbf{p}_{\mathrm{T}}^{\mathrm{lepton}})]})}
\end{equation}

\noindent Selection criteria on various observables define a signal region in which the signal to be searched for is enhanced and possible backgrounds suppressed. Two categories of backgrounds are distinguished. \textit{Irreducible} backgrounds yield the same or a very similar final state as the signal. \textit{Reducible} backgrounds show a different signature than the signal but might be misidentified due to e.g. mis-reconstruction effects or particles missed due to the detector acceptance. Different methods to estimate the background contribution in the signal regions are employed. 

In case of irreducible backgrounds, control regions are often used in which the background to be estimated is enhanced, while the signal contamination is negligible. A Monte Carlo (MC) simulation or template of the background process is compared to data measured and corrected accordingly. This correction is applied to the background MC or template in the signal region to obtain the final background estimate. The extrapolation is verified in validation regions which target a similar kinematic regime as the signal region but show an at most minor signal contamination.

Reducible backgrounds are often obtained directly from data by using data-driven methods like e.g. a matrix method. 
The estimation of backgrounds is summarized in Fig.~\ref{SUSY:analysis_flow}.

\begin{figure}
 \centerline{
  \includegraphics[width=0.9\textwidth]{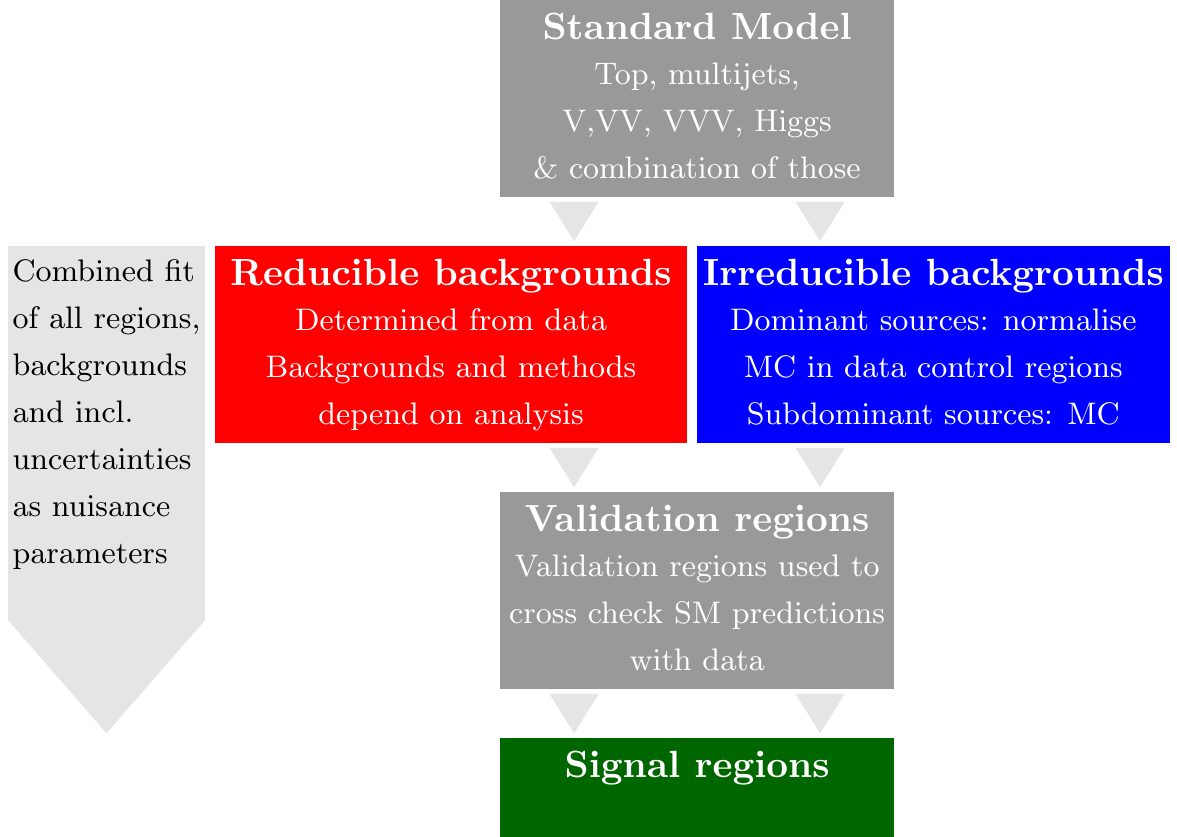}
  }
  \vspace*{8pt}
\caption{\label{SUSY:analysis_flow}The typical estimation of backgrounds in a search distinguishes between irreducible and reducible backgrounds. Depending on type and size, the background may be estimated using control regions or directly taken from data or MC simulation. Different backgrounds and uncertainties are combined in a simultaneous profile log likelihood fit. The background estimate obtained by this fit is first cross-checked in validation regions before being extrapolated to signal regions.}
\end{figure}

Backgrounds in all kinematic regions defined in an analysis are fitted to data in an simultaneous profile log likelihood fit including statistical, systematic and theoretical uncertainties.
By also including a signal model in this fit, exclusion limits on the signal can be derived or the size of an excess in data quantified.

While some analyses use simple selection criteria on kinematic quantities (\textit{cut-and-count} analyses), other fit distributions in observables particularly sensitive to the signal (\textit{shape} or \textit{multi-bin} analysis) or employ complex selection criteria using machine learning algorithms. 

\section{Searches for supersymmetric particles}
\label{SUSYsearches}

For a given mass, the production cross sections of gluinos and squarks of the first two generations is the largest at LHC as shown in Ref.~\refcite{susy_XSec_paper} and Fig.~\ref{susy_XSec}. The production proceeds via $pp \rightarrow \tilde{g}\tilde{g}$, $pp \rightarrow \tilde{q}\tilde{g}$, $pp \rightarrow \tilde{q}\tilde{q}$ and $pp \rightarrow \tilde{q}\tilde{q}^{*}$. Third generation squarks (stops and sbottoms) show smaller cross sections for comparable masses, and the cross sections for chargino or neutralino production are again smaller. However, if squarks and gluinos turn out to have too large masses to be in reach of the LHC, but charginos and neutralinos are relatively light, the production of charginos and neutralinos might still dominate. Cross sections for charginos and neutralinos depend on their wino/higgsino fraction and on if sleptons are lighter or not.

\begin{figure}
\centerline{
\includegraphics[width=0.7\textwidth]{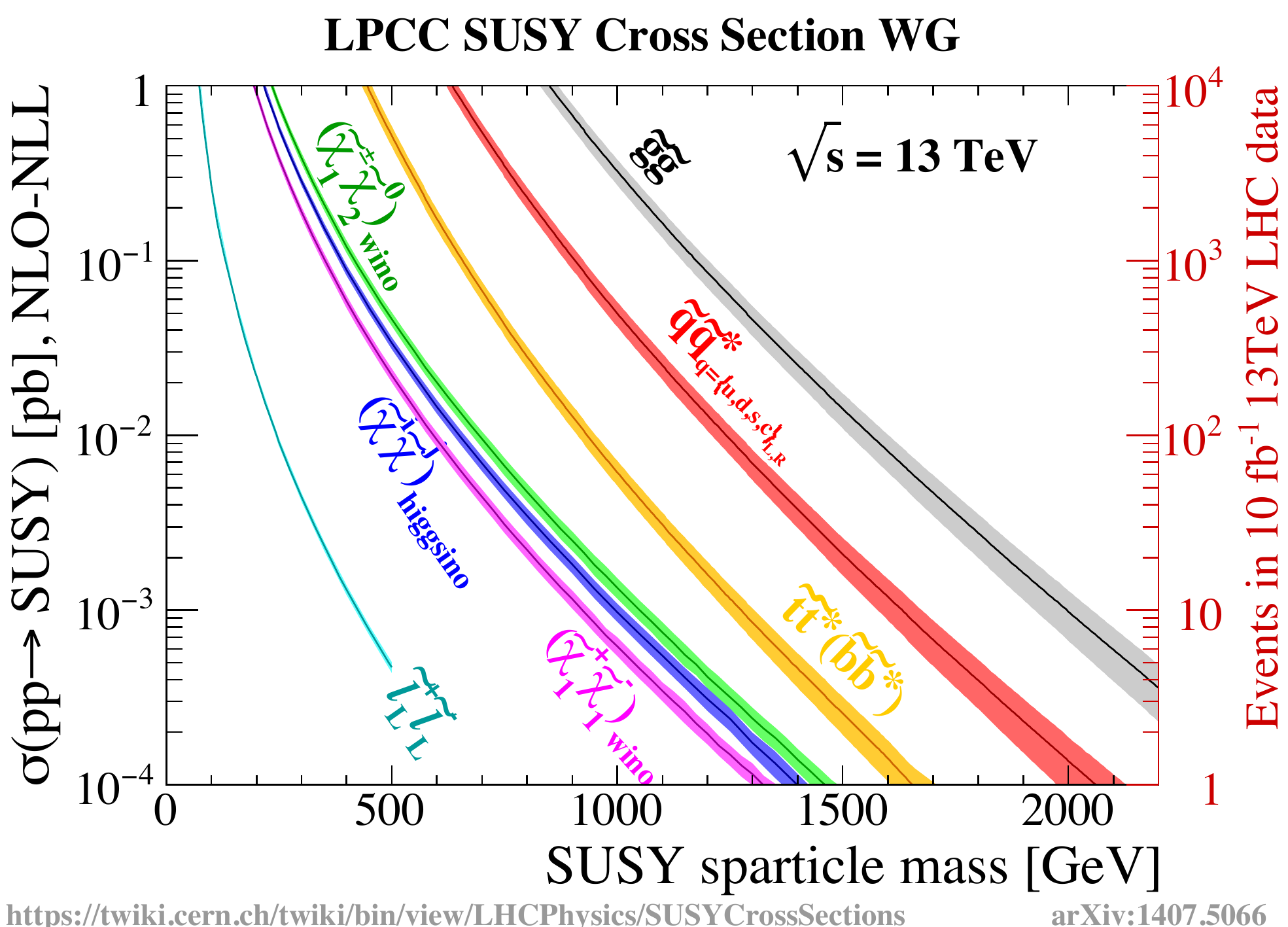}
}
\vspace*{8pt}
\caption{\label{susy_XSec} Production cross sections of supersymmetric particles at $\sqrt{s}= 13 \TeV$\cite{susy_XSec_paper,Beenakker:1996ch,Beenakker:1997ut}.}
\end{figure}

Due to smaller cross sections for charginos and neutralinos, searches for them particularly profit from the higher data statistics available using the full dataset of Run-2.

In the following, this review will highlight some exemplary analyses and techniques, focusing primarily on searches for supersymmetric particles.
Searches originally designed for SUSY models are sometimes interpreted in mediator dark matter simplified models as an example in Sec.~\ref{stop} shows. Vice-versa, generic BSM mediator searches can offer constraints on SUSY particles, as illustrated in Sec.~\ref{monoX}. All results by the ATLAS and CMS Collaborations may be found at their public pages\cite{ATLAS_SUSY,CMS_SUSY}.

\subsection{Searches for gluinos and squarks}
\label{strong}

Searches for gluinos and squarks of the first two generations quickly reached strong sensitivity during Run-2, as the cross-sections are relatively high. Thus, many strong-production analyses were already presented with a partial data set, and analyses using the full Run-2 statistics often show only modest improvements with respect to earlier results. Still, representative for many other searches, one search shall be presented here, which serves as `work-horse' due to its wide applications in reinterpretations. 

The \textit{0-lepton MHT} search\cite{CMS_0lepton} targets signatures with many jets (which may include jets containing heavy-flavor hadrons (\textit{b-tagged} jets)) and \met, vetoing the presence of leptons in the final state, and uses the complete Run-2 dataset of 137 \ifb\ collected by the CMS detector. Similar searches exist by ATLAS\cite{0lepton,ATLAS_0lepton_fullRun2,ATLAS_strong3,ATLAS_strong8}. Discrimination between background and signal is reached by a four-dimensional signal region with exclusive intervals (bins) in every dimension. The kinematic quantities used to define these dimensions are the jet multiplicity, the multiplicity of b-tagged jets, the sum of all transverse momenta of the jets selected and \met. In total 174 exclusive signal regions are defined and ensure to capture the characteristics of different signal hypotheses. Important backgrounds include \ttbar, single top and  $W/Z$+jets processes. No significant excess is observed in any of the signal regions. Exclusion limits at 95 \% CL are derived by combination of the search regions in a likelihood fit (\textit{multi-bin} fit). Fig~\ref{strong:0lepton} shows exemplary limits on pair production of gluinos with assumed decays $\tilde{g} \rightarrow q\bar{q} \tilde{\chi}_{1}^{0}$ and on pair production of squarks with $\tilde{q} \rightarrow q \tilde{\chi}_{1}^{0}$, but limits were set on third-generation squarks as well. In case of gluinos, the limits may reach as high as 2 \TeV\ for very small \ninoone\ masses, with even stronger limits reached in other simplified models. However, the limits weaken substantially if considering sizable \ninoone\ masses and are only valid for the specific simplified model under consideration. In case of the limit on $\tilde{q} \rightarrow q \tilde{\chi}_{1}^{0}$ two different limit curves are derived. The curve yielding stronger limits assumes four degenerate squark flavors for the superpartners of the up, down, charm and strange quarks; the limits reach up to 1630 \GeV\ at most. Reduced limits of 1130 \GeV\ are obtained if only assuming one squark type to be in reach of the LHC while all other squarks would be too heavy to contribute. A preliminary result\cite{ATLAS_0lepton_fullRun2} by the ATLAS Collaboration reaches stronger limits on gluino and squark masses by a few hundreds of \GeV, while yielding similar or less strong limits on the LSP masses. The \textit{0-lepton MHT} search also includes signal regions requiring the presence of only one jet and \met (an equivalent ATLAS search is Ref.~\refcite{Aaboud:2017phn}). These regions are able to address ultra-compressed scenarios with very small $\Delta m(\tilde{g},\ninoone)$, resulting in very low-energetic decay products from SUSY particles.

\begin{figure}
 \centerline{
  \includegraphics[width=0.49\textwidth]{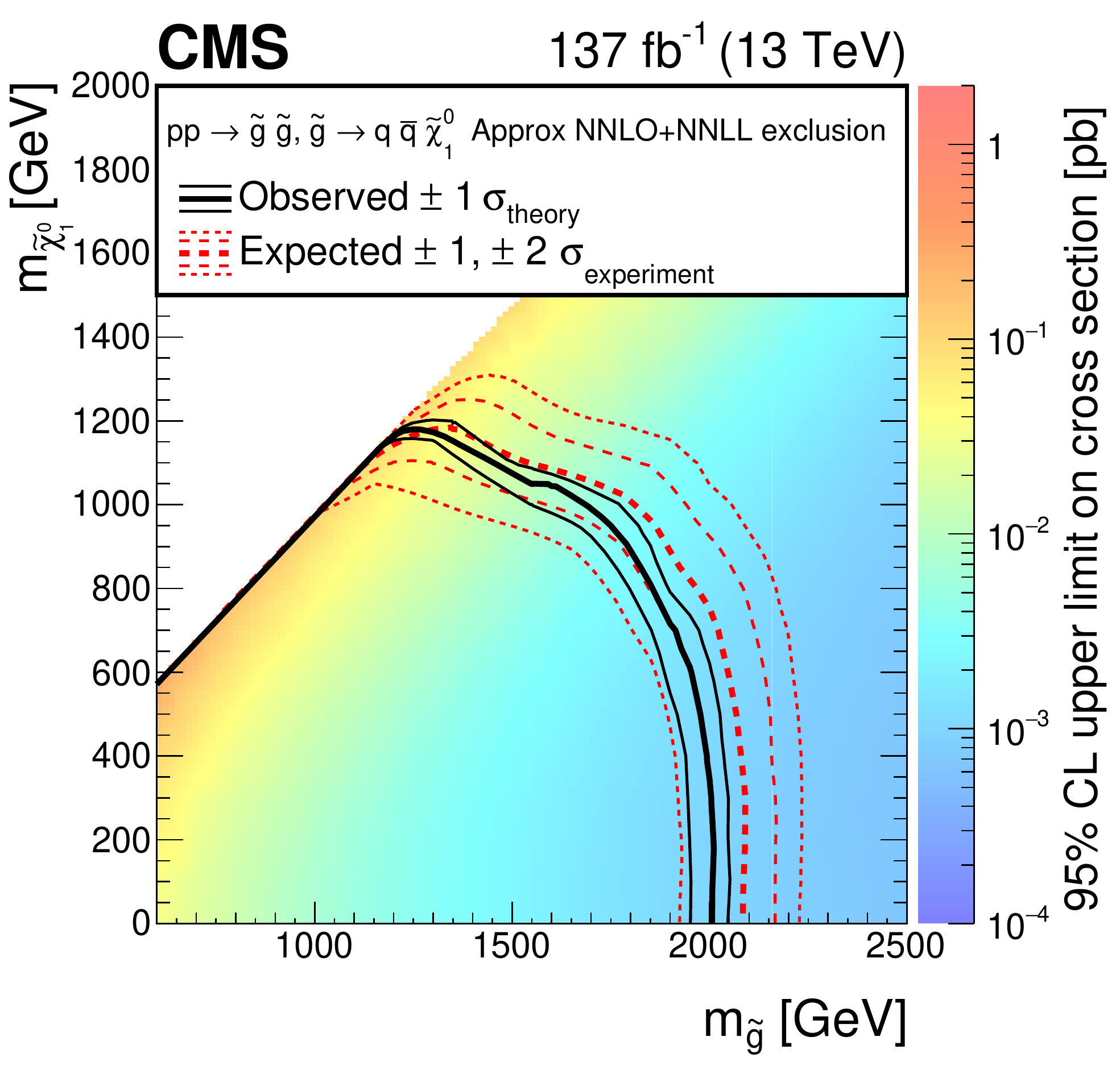}
  \includegraphics[width=0.49\textwidth]{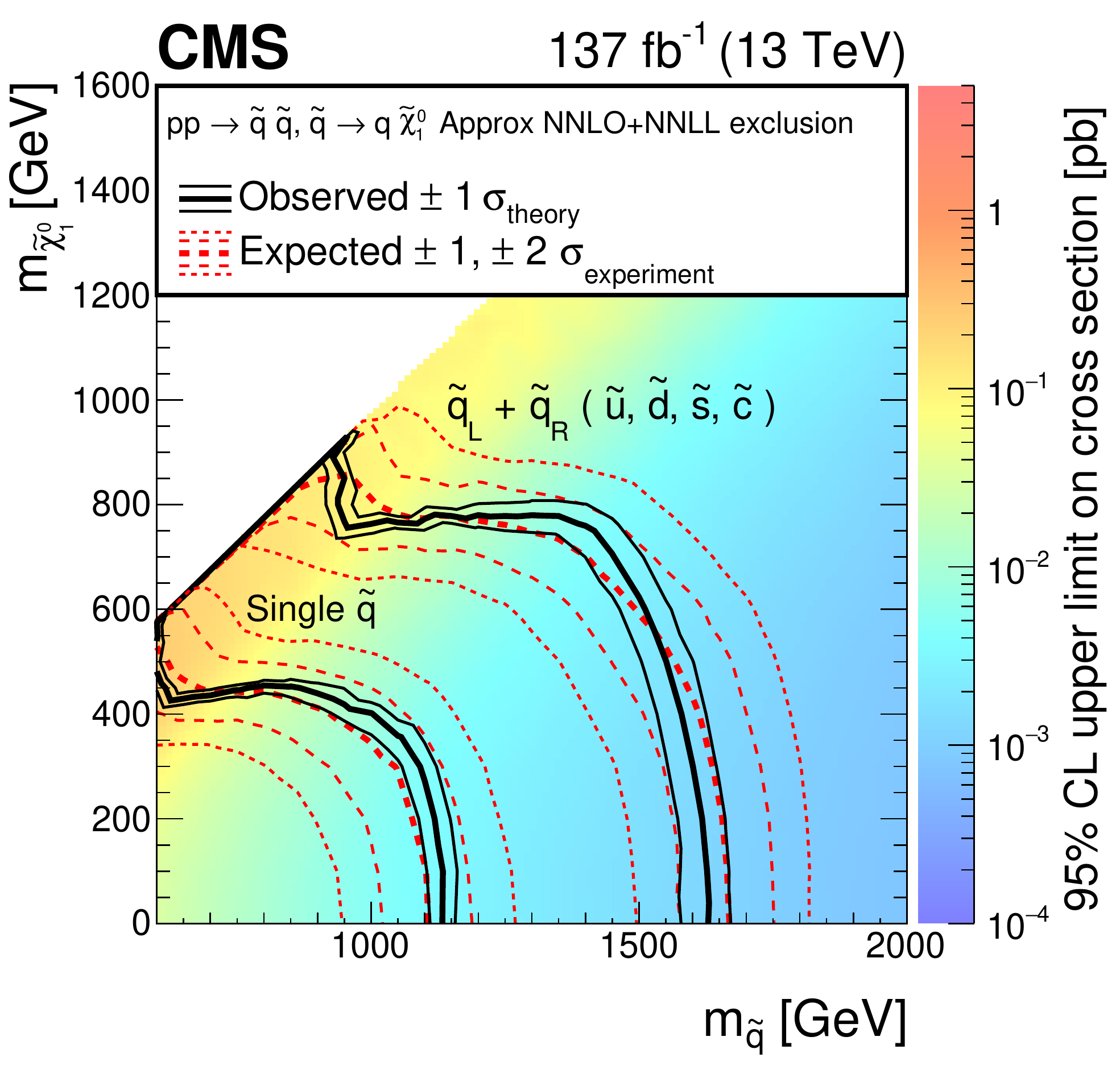}  
 }
 \vspace*{8pt}
\caption{\label{strong:0lepton}Exclusion limits\cite{CMS_0lepton} at 95\% CL considering a simplified model with pair production of gluinos and $\tilde{g} \rightarrow q\bar{q} \tilde{\chi}_{1}^{0}$ (left) and with pair production of squarks and $\tilde{q} \rightarrow q \tilde{\chi}_{1}^{0}$ (right).}
\end{figure}

Searches for gluinos and squarks are also performed in signatures including charged leptons or photons\cite{ATLAS_strong1,ATLAS_strong2,ATLAS_strong3,ATLAS_strong4,ATLAS_strong5,ATLAS_strong6,ATLAS_strong7,CMS_strong1,CMS_strong2}.
No significant excess has been seen until now. Exclusion limits reach similar sizes as for the \textit{0-lepton MHT} search depending on the model considered.

\subsection{Searches for stop and sbottom quarks}
\label{stop}

Searches for stop and sbottom quarks take a special role, as \textit{natural} SUSY models require these to be not too heavy due to a strong coupling to the Higgs as described in Sec.~\ref{SUSY}. 
In this, they differ from other squarks which might be heavy without causing large fine-tuning to compensate radiative corrections to the Higgs mass. 

The decay of the lighter stop, $\tilde{t}_{1}$, depends on the mass difference between $\tilde{t}_{1}$ and \ninoone\ as depicted in Fig.~\ref{stop_decays}. If the mass difference is larger than the mass of the top, a \textit{two-body} decay occurs: $\tilde{t}_{1} \rightarrow t\ninoone$. If the mass difference is smaller than the mass of the top, but still larger than the mass of the $W$ boson, a \textit{three-body} decay appears: $\tilde{t}_{1} \rightarrow Wb\ninoone$. In case of even smaller mass differences, a \textit{four-body} decay happens: $\tilde{t}_{1} \rightarrow ff'b\ninoone$ where $ff'$ may be a lepton-neutrino pair or a quark pair. 

\begin{figure}
 \centerline{
  \includegraphics[width=0.7\textwidth]{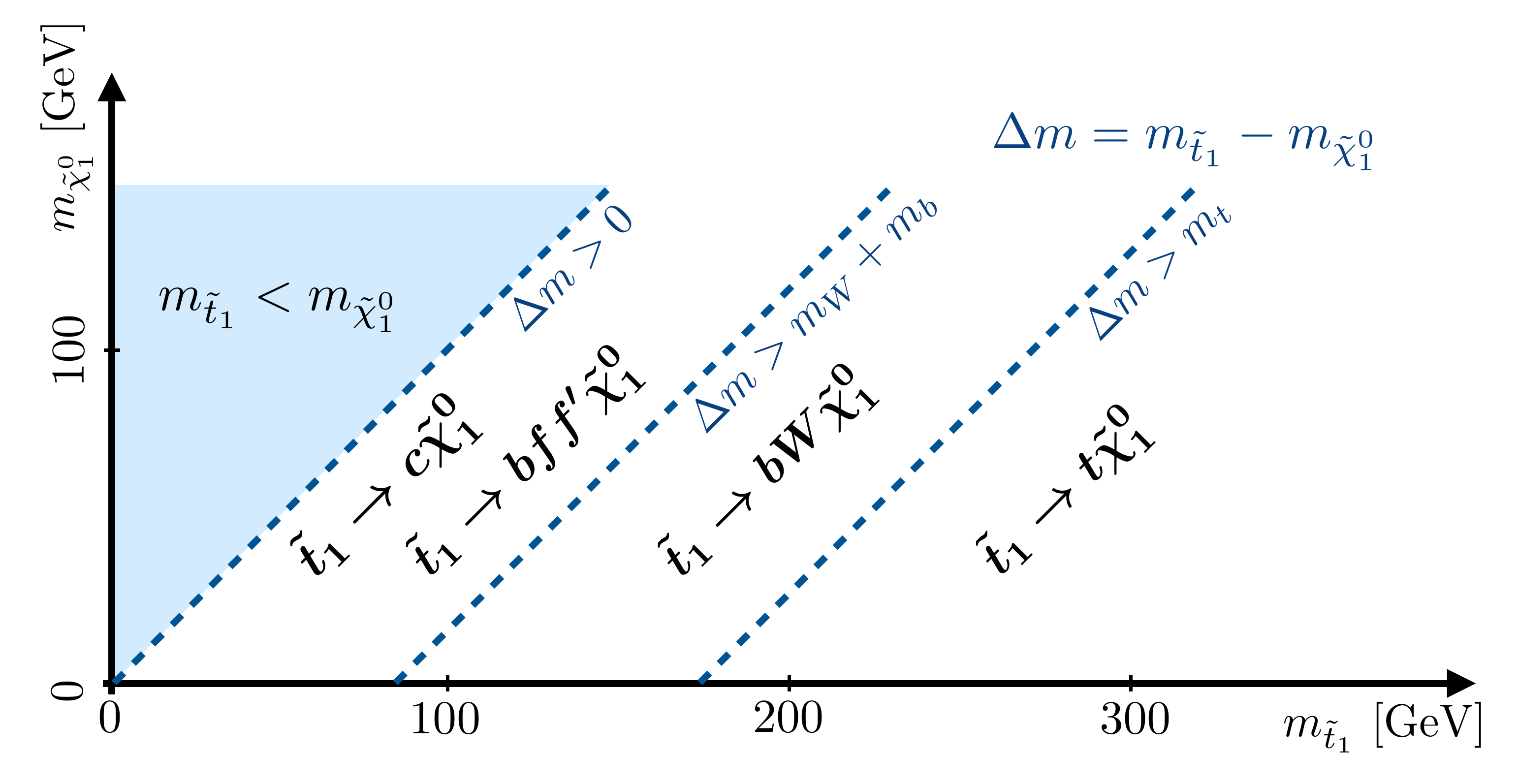}
 }
 \vspace*{8pt}
\caption{\label{stop_decays}Different possibilities of stop quark decays depending on the available mass difference $\Delta m(\tilde{t}_{1},\ninoone)$. The decay may proceed as two-, three- or four-body decay\cite{stop_1L}.}
\end{figure}

Searches for stop quarks are conducted in different signatures, including no leptons or (multiple) leptons~\cite{CMS_0lepton,stop_1L,ATLAS_stop_3body,ATLAS_stop2,ATLAS_stop3,ATLAS_stop4,CMS_stop1,CMS_stop2,CMS_stop3,CMS_stop4,CMS_stop4,CMS_stop5,CMS_stop6}.

In case of the three-body decay, a recent preliminary analysis\cite{ATLAS_stop_3body} by the ATLAS Collaboration considering signatures with an isolated low-energetic electron or muon was able to derive exclusion limits up to 720 \GeV\ on the $\tilde{t}_{1}$ mass for \ninoone\ masses up to 580 \GeV\ for the region $m(W) + m(b) \leq \Delta m(\tilde{t},\ninoone) \leq m(t)$. The search analyzed the full dataset of 139 \ifb\ of Run-2. A recurrent neural network together with a shallow neural network is used to obtain sensitivity to this challenging kinematic region resembling the production of \ttbar. Angles, transverse momenta, and (missing) energies of jets and the lepton are inputs to the networks. The output distribution is further subdivided in exclusive regions to be combined in a likelihood fit. The analysis is able to reach sensitivity beyond the design region, reaching into the kinematic regions of the two- and four-body decays, as shown in Fig.~\ref{stop_summary}. This figure summarizes constraints on $m(\tilde{t}_{1})$ and $m(\ninoone)$ by different searches, including 
the \textit{jet+\met} analysis\cite{Aaboud:2017phn} mentioned in Sec.~\ref{strong}. This analysis is able to exclude stop masses up to 390 \GeV\ in the four-body region for the case $m(\tilde{t}_{1}) - m(\ninoone) \sim m(b)$.

\begin{figure}
 \centerline{
  \includegraphics[width=0.9\textwidth]{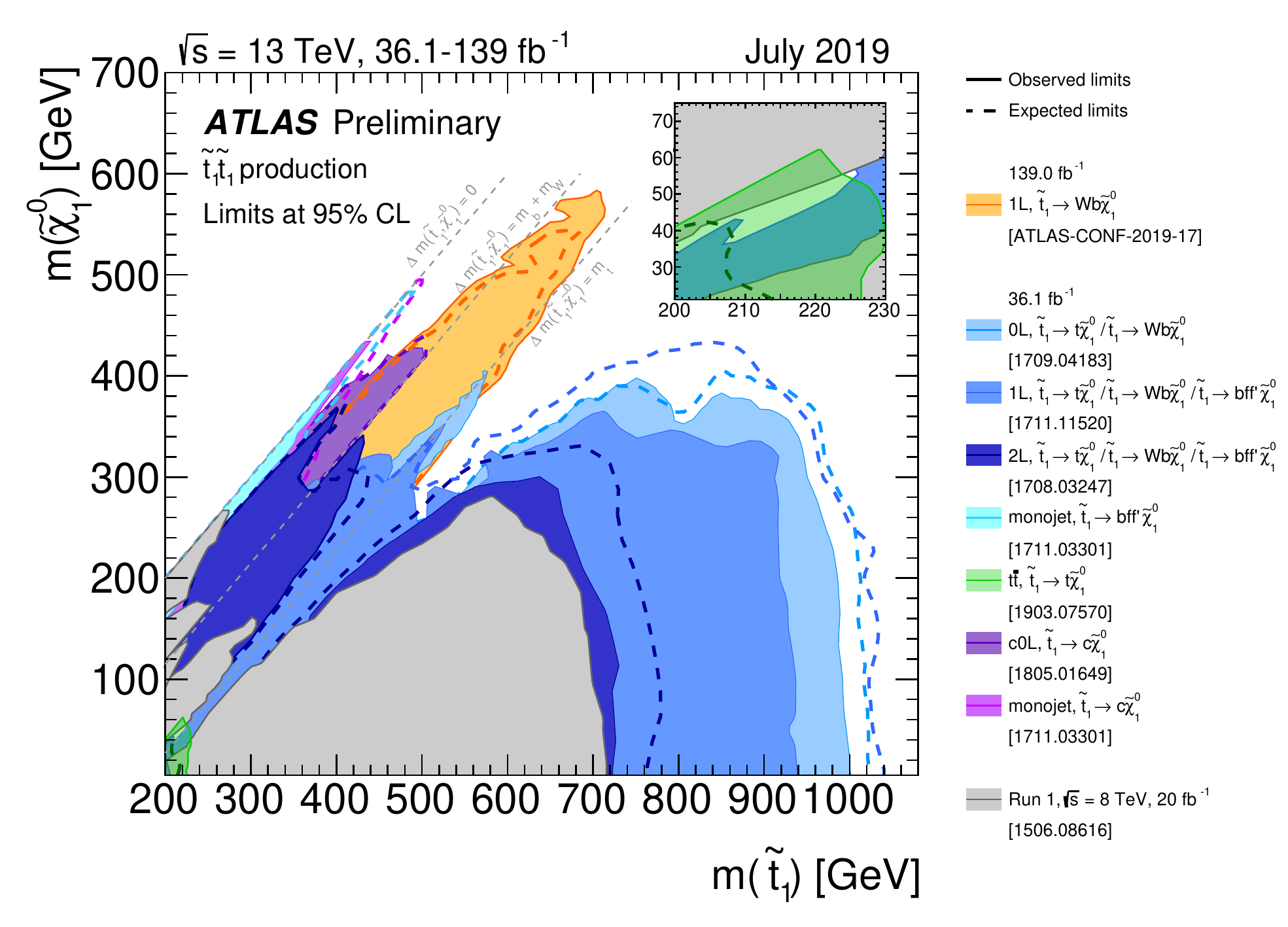}
 }
 \vspace*{8pt}
\caption{\label{stop_summary}Summary of ATLAS searches for the lighter stop quark. Also kinematic difficult regions in which $\Delta m = m(\mathrm{top})$ or $\Delta m = m(W) + m(b)$ are more and more addressed\cite{stop_summary_plot}}.
\end{figure}

Searches for stop and sbottom quarks in signatures without or with a lepton\cite{stop_1L,ATLAS_stop2} can be interpreted (using 36.1 \ifb) in a specific slice of the pMSSM with sbottom/stop production and with the parameters set to yield the correct relic density and Higgs mass, see Fig.~\ref{welltempered}. In this \textit{well-tempered neutralino} model the masses of the three lightest neutralinos and the lightest chargino are within 50 \GeV. All different decay modes of sbottom and stop quarks are considered, as well as different mixing assumptions. Exclusion limits are weaker for $\tilde{t}_{1} \approx \tilde{t}_{R}$ than for $\tilde{t}_{1} \approx \tilde{t}_{L}$.

\begin{figure}
 \centerline{
 \includegraphics[width=0.6\textwidth]{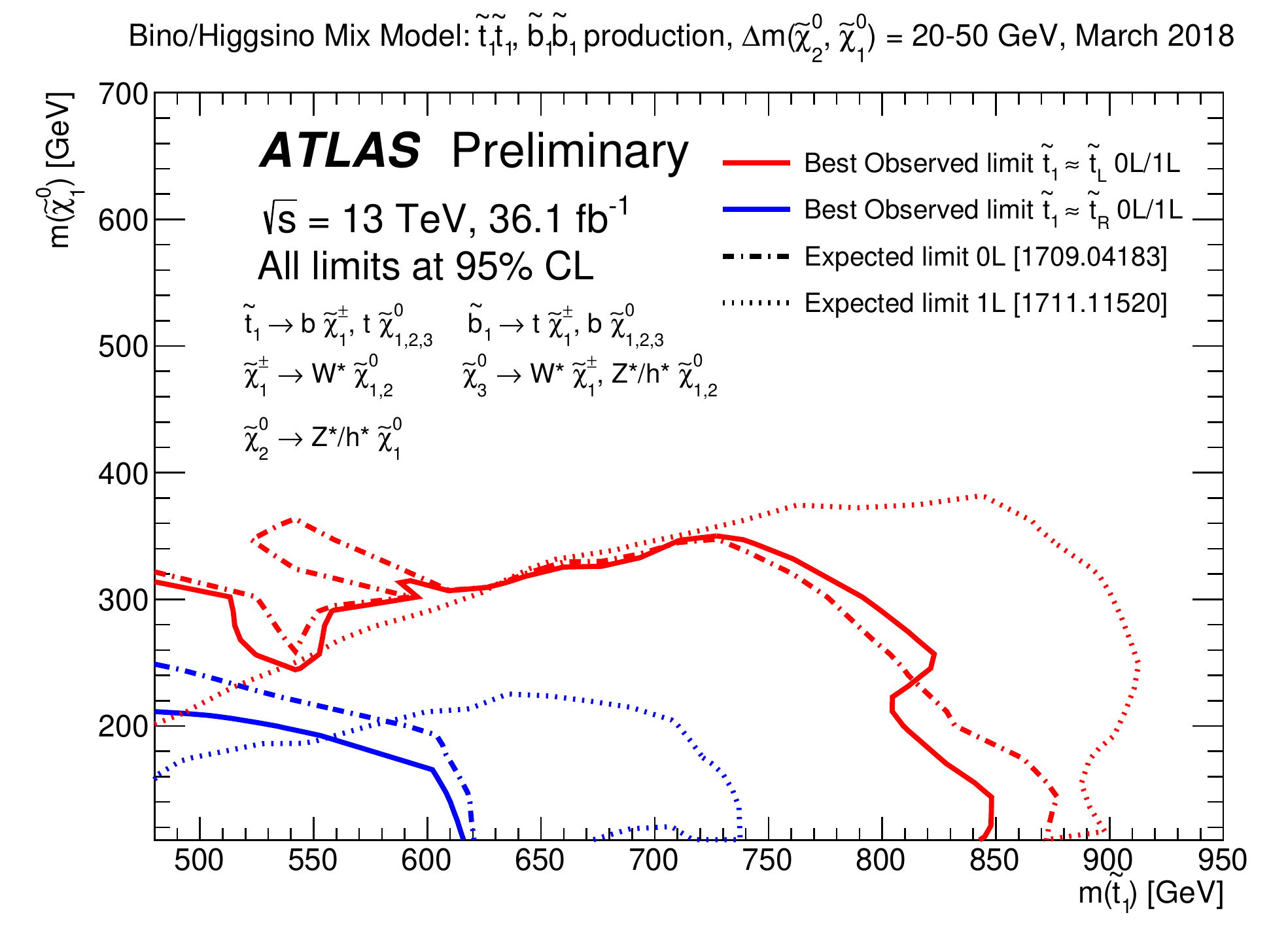}
 }
 \caption{\label{welltempered}Interpretation of searches for stop and sbottom quarks in signatures with and without lepton in the \textit{well-tempered neutralino} pMSSM model\cite{stop_summary_plot}.}
\end{figure}

Although originally designed as searches for stops and sbottoms, searches often feature signal regions designed for BSM mediator simplified DM models\cite{stop_1L,ATLAS_stop2,ATLAS_stop3,ATLAS_stop5}, specifically for spin-0 mediators.
The coupling of a spin-0 mediator to SM particles is constrained by precision flavor measurements, but these constraints are relaxed by introducing Minimal Flavor Violation\cite{MFV}. A consequence is that color-neutral spin-0 mediators might be produced in association with heavy-flavor quarks or via loop-induced gluon-fusion.
ATLAS searches for stops with signatures without or with one or two lepton(s) set limits on the mass of the (pseudo-)scalar mediator particle assuming a light DM particle of 1 \GeV\ and a common coupling of $g=1$ to SM and DM particles\cite{ATLAS_DMsummary}. The limits extend up to 45 \GeV\ for a scalar mediator and 15 -25 \GeV\ for a pseudo-scalar mediator.

\subsection{Searches for weakly produced supersymmetric particles}

The decay pattern of charginos and neutralinos (electroweakinos) into lighter neutralinos and charginos depends on the respective fraction of winos, binos and higgsinos, as they inherit the couplings of their SM partner $W$, $Z$ and Higgs bosons. Decays into (s)leptons and (s)quarks are also possible if kinematically allowed. The signature obtained depends on if $\mu$ is larger or smaller than $M_{1}$ and $M_{2}$. For $\mu < M_{1} < M_{2}$ the lightest SUSY particles \ninoone, \ninotwo\ and \chinoone\ form a higgsino triplet state with similar masses as shown in Fig.~\ref{WinoBinoHiggsino}. A light higgsino is suggested by naturalness arguments\cite{barbieri,carlos}. If $M_{1} < M_{2} \ll \mu$, the lighter charginos and neutralinos have a large wino and bino component.

\begin{figure}
 \centerline{
  \includegraphics[width=0.6\textwidth]{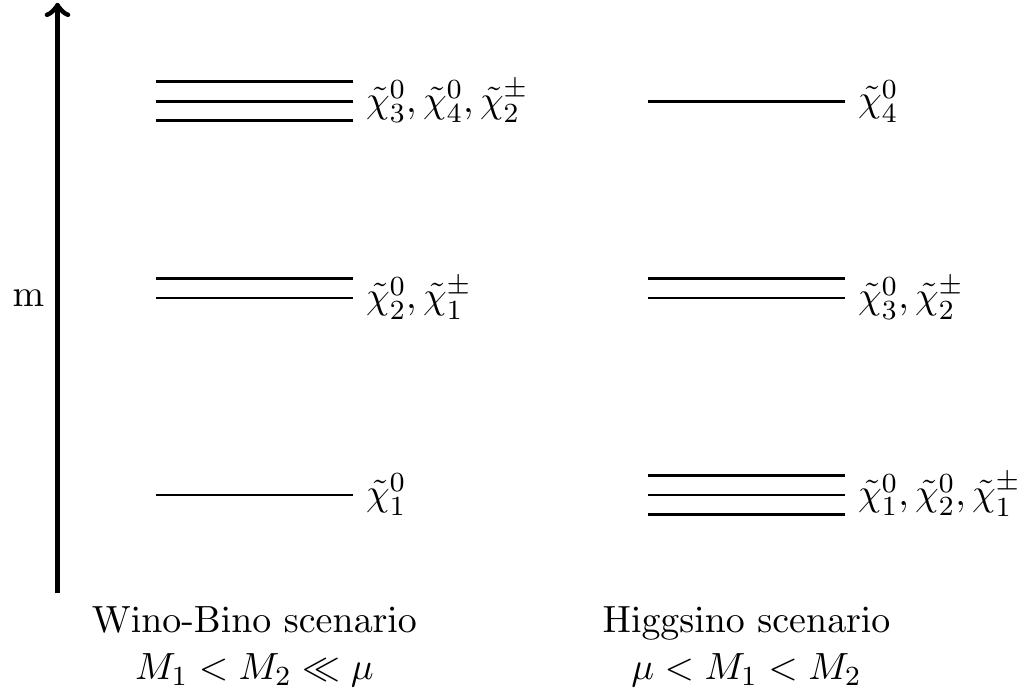}
 }\vspace*{8pt}
\caption{\label{WinoBinoHiggsino}Depending on the relative size of $M_1$, $M_2$ and $\mu$, the lighter charginos and neutralinos are close in mass and almost degenerate (for $\mu <M_1 < M_2$) and the LSP shows a large higgsino contribution, or the masses of the lighter charginos and neutralinos are notably lager than the mass of the LSP (for $M_1 < M_2 \ll \mu$). In this case the LSP shows a large bino contribution (and the NLSPs a wino contribution).}
\end{figure}

Leptons are often produced in decays of charginos and neutralinos, e.g. in decays of emitted gauge bosons. Many searches for electroweakinos thus focus on signatures with leptons. This allows a good discrimination against irreducible backgrounds mainly consisting of diboson and \ttbar+X processes. Backgrounds including tops can usually be well suppressed by vetoing the presence of b-tagged jets. However, recently, the first fully-hadronic analyses have been presented which require the presence of three or more b-tagged jets and achieve background suppression through the good abilities of the experiments to identify b-tagged jets.

\subsubsection{Searches for higgsinos}

\begin{figure}
 \centerline{
  \includegraphics[width=0.9\textwidth]{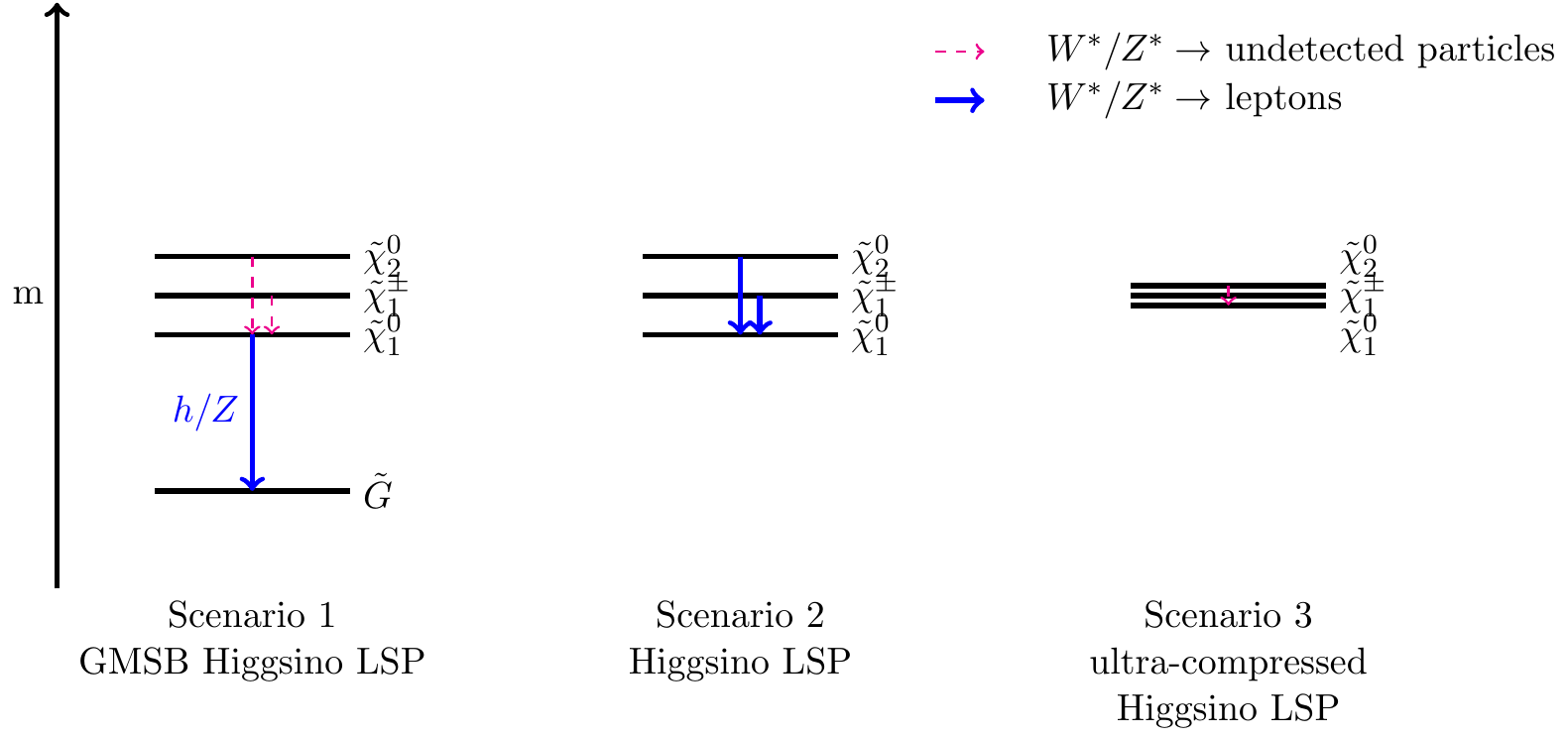}
 }
 \vspace*{8pt}
\caption{\label{higgsino_scenarios}Different scenarios for decays of Higgsinos.}
\end{figure}

Searches for higginos are challenging due to low-energetic decay products. Fig.~\ref{higgsino_scenarios} summarizes the different scenarios possible in searches for higgsinos. In cases with a \textit{GMSB Higgsino NLSP} in SUSY models with gauge-mediated symmetry breaking (GMSB)\cite{GMSB1,GMSB2} the masses of \chinoone, \ninotwo\ and \ninoone\ are close together. The \ninoone\ is the NLSP and decays to a gravitino $\tilde{G}$. Particles ($h$ or $Z$) with enough transverse momentum to be detected are only emitted in the decay $\ninoone \rightarrow h/Z \tilde{G}$. Thus the signature in the detector consists entirely of the decays products of the emitted $h$ or $Z$ bosons, while any other emitted particles of the electroweakino decays are too low-energetic to be detected. 
In the second scenario a \textit{higgsino LSP} with some small bino or wino component is assumed, leading to a compressed mass spectra between \chinoone, \ninotwo\ and \ninoone, where the \ninoone\ is the LSP. Only low-momentum particles are emitted in the decays, which are however visible in the detector if being able to control the reconstruction of particles of low momenta. This is in contrast to the third scenario in which the mass differences is in the order of $\sim 100 \MeV$. Only extremely low-energetic particles emerge from this \textit{ultra-compressed higgsino LSP} scenario, causing the common particle reconstruction algorithm to fail. Instead, distinctive signatures of disappearing tracks are obtained as illustrated below.

The \textit{GMSB Higgsino NLSP} scenario is addressed by an ATLAS search\cite{ATLAS_fourb} considering signatures with three or four b-tagged jets in addition to \met, using 36.1 \ifb. These are obtained from the decays of the two Higgs bosons and are an excellent handle to suppress SM backgrounds. Due to this the search is able to consider final states without leptons and thus profits from high branching ratios. Two sets of SRs are considered with the first set requiring low \met\ targeting SUSY models with low $\mu$, while the other requiring high \met\ targets scenarios with high $\mu$. No significant excess is seen. Searches considering the \textit{GMSB Higgsino NLSP} scenario are also covered by CMS\cite{CMS_higgsino1,CMS_higgsino2,CMS_higgsino3}.

Searching for the \textit{higgsino LSP} in compressed scenarios is possible due to the excellent performance of ATLAS and CMS to reconstruct low-\pt\ electrons and muons\cite{ATLAS_higgsino,CMS_higgsino4}. In case of the ATLAS analysis\cite{ATLAS_higgsino} (using 36.1 \ifb) electrons with a \pt\ down to 4.5 \GeV\ and muons down to 4 \GeV\ are reconstructed as well as an invariant mass of the two leptons down to $m_{ll} = 1 \GeV$. In addition to the two leptons, a jet steaming from initial-state-radiation is also required (as shown in Fig.~\ref{higgsino_searches}) which imbalances the whole system and thus results in higher \met. Fig.~\ref{higgsino_limits} illustrates the necessity to reconstruct low $m_{ll}$ values, as an higgsino LSP leads to a $m_{ll}$ peak at significantly lower values than in case of a wino or bino LSP. Reconstructing low-energetic leptons however increases the background with mis-identified leptons considerably, making thus a precise estimation of this background essential. No significant excess has been seen, but this analysis allowed to surpass the LEP constraints on the \chinoone\ for the first time at the LHC. Exclusion limits at 95\% CL are displayed in Fig.~\ref{higgsino_limits} together with limits from an analysis searching for disappearing tracks\cite{ATLAS_distrack}, targeting \textit{ultra-compressed higgsinos LSP} scenarios. In this case, a chargino with a long life-time decays to an invisible \ninoone\ and a low-energetic pion which is not able to pass the pixel tracking system as illustrated in Fig.~\ref{higgsino_searches}. Reconstruction of the resulting short tracks by the disappearing chargino, so-called \textit{pixel-only tracklets}, are only possible due to the insertion of the Insertable-b-Layer (IBL) between Run 1 and Run 2 which allows reconstruction of track lengths down to 12 cm at ATLAS.

\begin{figure}
 \centerline{
  \includegraphics[width=0.26\textwidth]{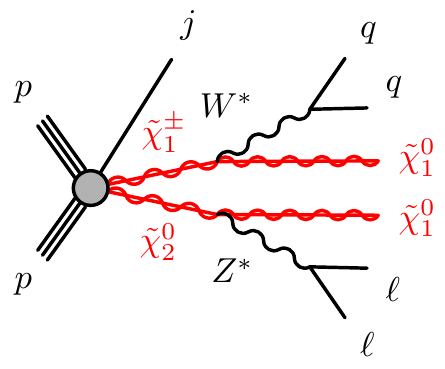}
  \includegraphics[width=0.21\textwidth]{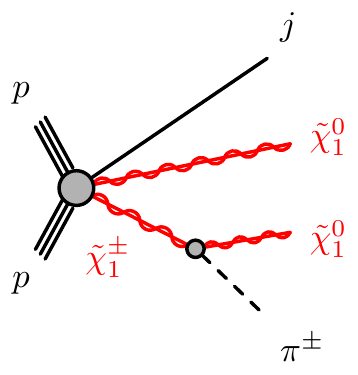} 
  \includegraphics[width=0.5\textwidth]{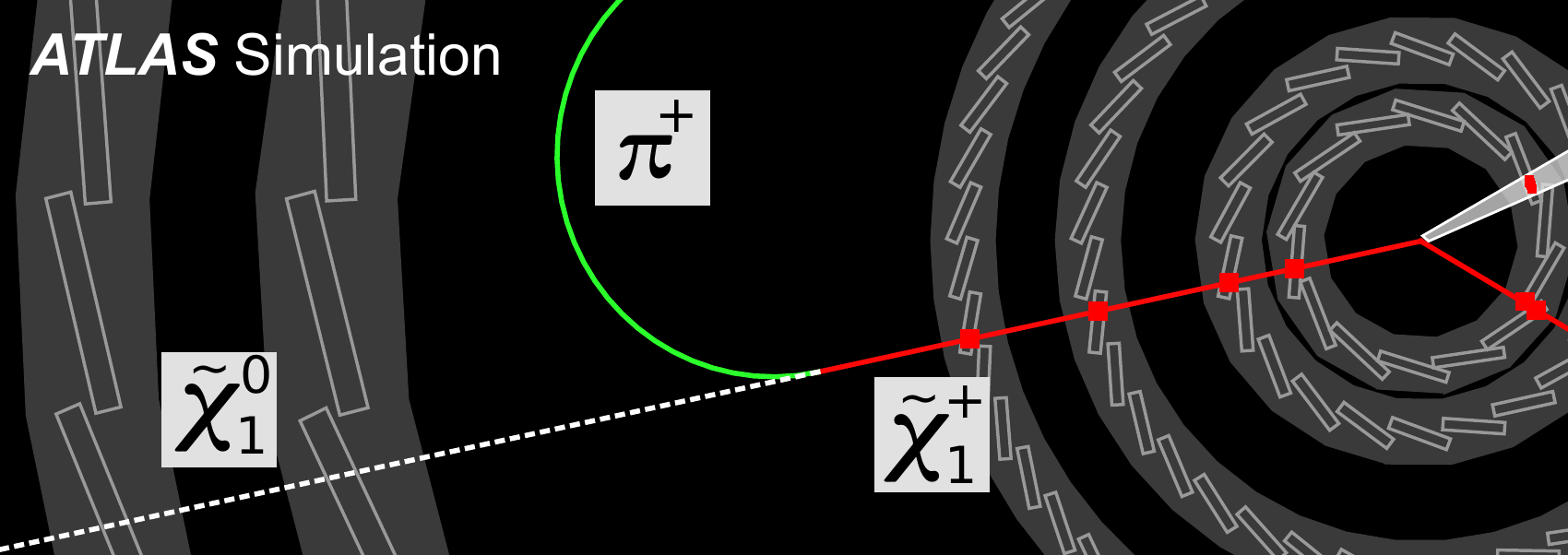}   
 }\vspace*{8pt}
\caption{\label{higgsino_searches}Decay diagram for the \textit{higgsino LSP}\cite{ATLAS_higgsino} (left) and the \textit{ultra-compressed higgsino LSP} (middle) scenarios. The signature in the ATLAS detector for the \textit{ultra-compressed higgsino LSP} scenario\cite{ATLAS_distrack} (right).}
\end{figure}

\begin{figure}
 \centerline{
  \includegraphics[width=0.49\textwidth]{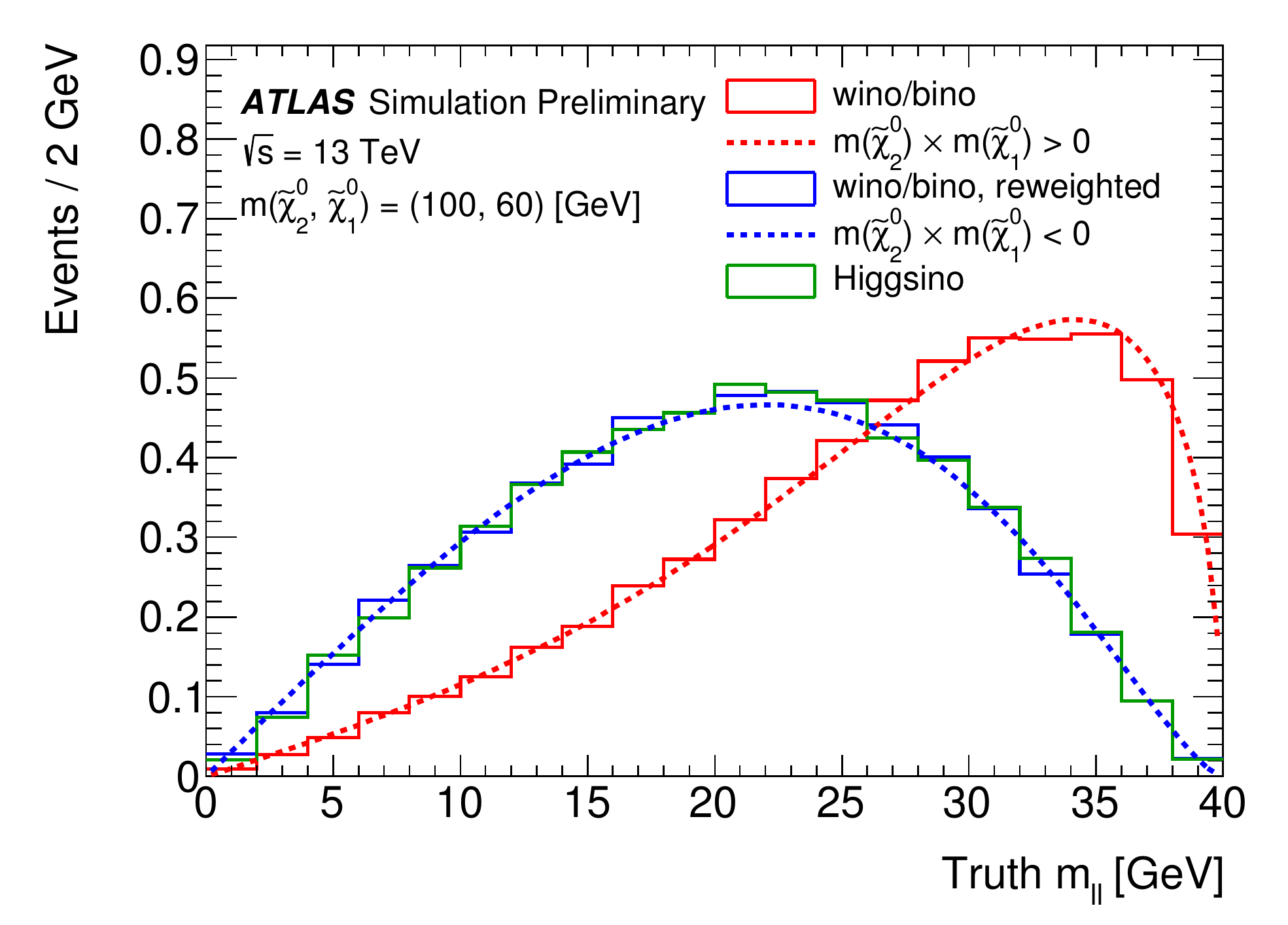}
  \includegraphics[width=0.49\textwidth]{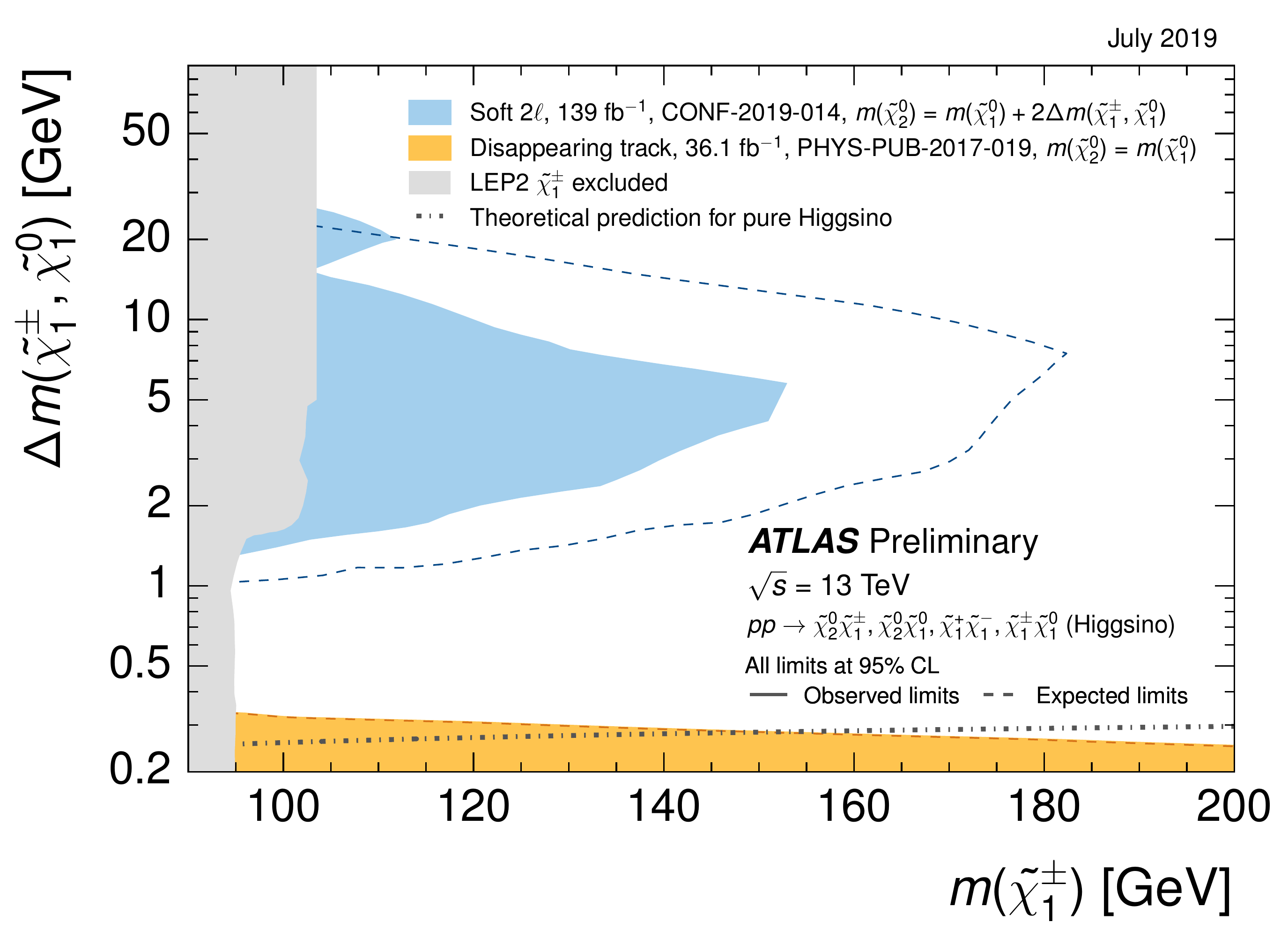}
 }
 \vspace*{8pt}
 \caption{\label{higgsino_limits}The invariant mass of both leptons for higgsino or wino/bino LSPs\cite{ATLAS_higgsino_CONF} (left). Exclusion limits at 95\% CL of the higgsino search and the disappearing track search\cite{stop_summary_plot} (right).}
\end{figure}

\subsubsection{Searches for winos/binos}

In contrast to searches for an higgsino LSP, searches for winos and binos result in a larger mass difference $\Delta m$ between charginos and neutralinos and consequently in emission of higher energetic particles as shown in Fig.~\ref{WinoBinoHiggsino}. The LSP is assumed to be bino-like and a singlet state. The NLSPs are often assumed to be wino-like. Production of winos profit from higher cross-sections compared to higgsino production as shown in Fig.~\ref{susy_XSec}. Searches for two or three leptons in the final state address many different decay possibilities of charginos, neutralinos or sleptons as illustrated in Fig~\ref{23L_diagrams}. The leptons originate from $W$- or $Z$ bosons decays or directly from slepton decays. The ATLAS \textit{2/3-lepton} search\cite{23L_conventional} addresses the different signatures expected by a set of three signal regions. The first set of SRs requiring two electrons or muons and no jets targets models with direct or indirect production of sleptons like depicted in Fig.~\ref{23L_diagrams} in the left and middle left diagram. The second set focuses on scenarios leading to two leptons and at least two jets originating from gauge bosons produced in the decays of charginos and neutralinos (Fig.~\ref{23L_diagrams} middle right). Signal regions requiring three leptons consider similar scenarios (Fig.~\ref{23L_diagrams} right).

\begin{figure}
 \centerline{
  \includegraphics[width=0.24\textwidth]{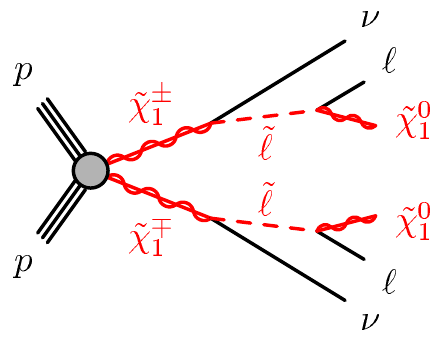}
  \includegraphics[width=0.24\textwidth]{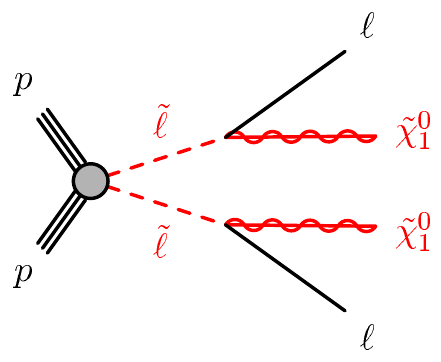}
  \includegraphics[width=0.24\textwidth]{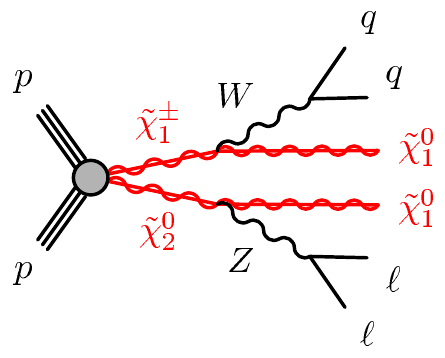}
  \includegraphics[width=0.24\textwidth]{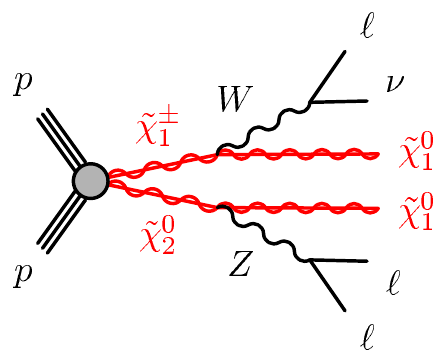}  
 }
 \vspace*{8pt}
\caption{\label{23L_diagrams}Possible decays of pair-produced \chinoone\ and \ninotwo\ into final states with two or three leptons\cite{23L_conventional}.}
\end{figure}

Different kinematic observables are used to discriminate signal from background such as \met, the invariant mass of both leptons $m_{ll}$ or the transverse momenta of the lowest energetic lepton, but the most important variable is \mttwo\cite{Barr:2009jv}: 

\begin{equation}
 \mttwo = \min_{\mathbf{q}_{\mathrm{T}}} [\max{(m_{\mathrm{T}}(\mathbf{p}_{\mathrm{T}}^{l_1},\mathbf{q}_{\mathrm{T}}),m_{\mathrm{T}}(\mathbf{p}_{\mathrm{T}}^{l_2},\mathbf{p}_{\mathrm{T}}^{\mathrm{miss}} - \mathbf{q}_{\mathrm{T}}))}]
\end{equation}

\noindent where $\mathbf{p}_{\mathrm{T}}^{l_1}$ and $\mathbf{p}_{\mathrm{T}}^{l_2}$ are the transverse momenta of both leptons and $\mathbf{q}_{\mathrm{T}}$ minimizes the larger of $m_{\mathrm{T}}(\mathbf{p}_{\mathrm{T}}^{l_1},\mathbf{q}_{\mathrm{T}})$ and $m_{\mathrm{T}}(\mathbf{p}_{\mathrm{T}}^{l_2},\mathbf{p}_{\mathrm{T}}^{\mathrm{miss}} - \mathbf{q}_{\mathrm{T}})$ with $m_{\mathrm{T}}(\mathbf{p}_{\mathrm{T}}^{l_1},\mathbf{q}_{\mathrm{T}}) = \sqrt{2 (p_{\mathrm{T}}q_{\mathrm{T}} - \mathbf{p}_{\mathrm{T}} \cdot \mathbf{q}_{\mathrm{T}})}$. Backgrounds such as \ttbar\ and $WW$ are suppressed using this variable, as \mttwo\ is bounded from above by the mass of the $W$ boson for backgrounds with $W \rightarrow l\nu$ in contrast to the SUSY signal yielding larger values. Distributions of signal and backgrounds are shown in Fig.~\ref{mt2_plot} for a signal region requiring two leptons. None of the signal regions in the search reports a significant excess. An example exclusion limit is shown in Fig.~\ref{mt2_plot} for a simplified model assuming chargino/neutralino production with decays mediated by $W/Z$ bosons.

\begin{figure}
 \centerline{
  \includegraphics[width=0.47\textwidth]{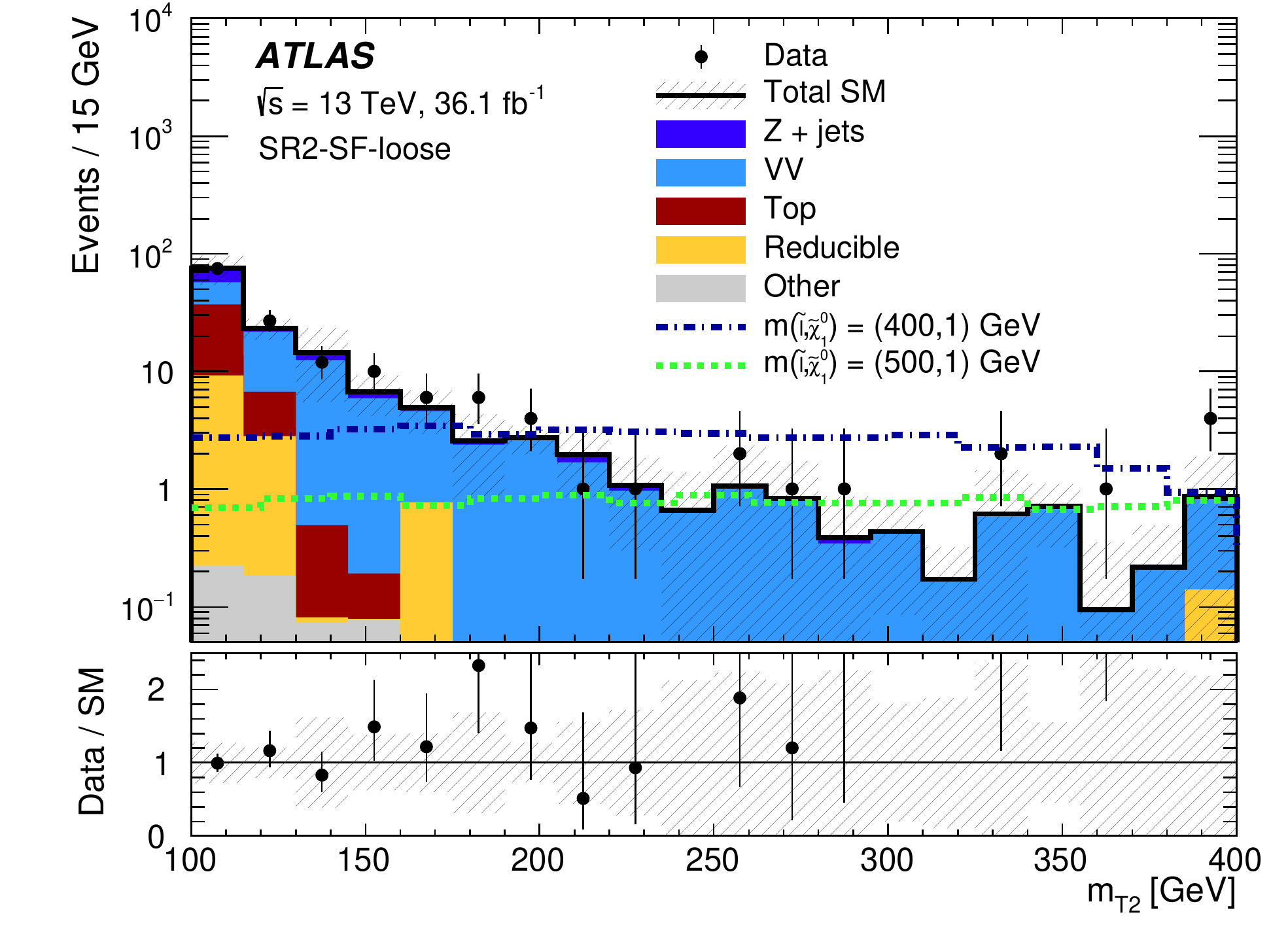}
  \includegraphics[width=0.52\textwidth]{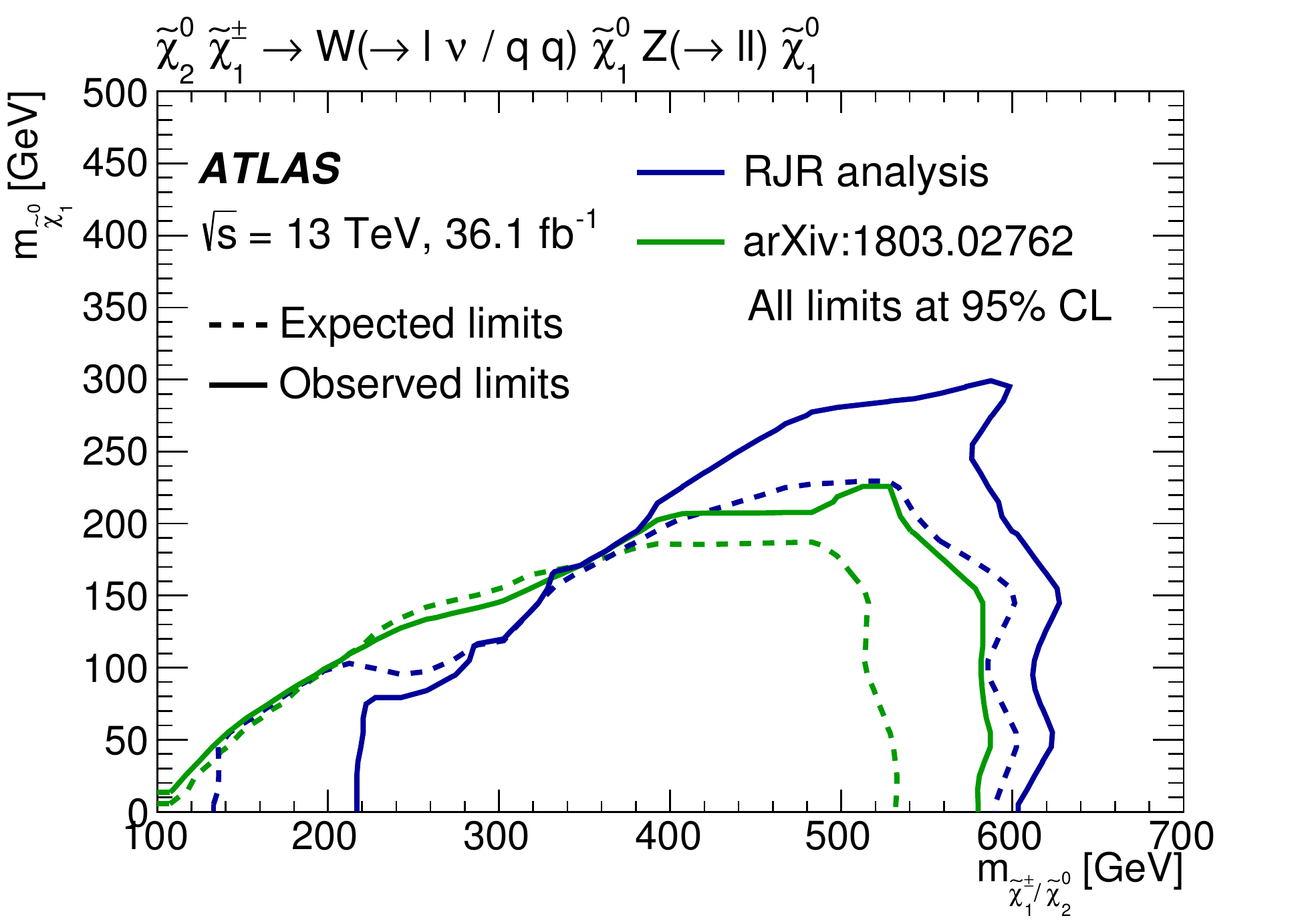}  
 }
 \vspace*{8pt}
\caption{\label{mt2_plot}Suppression of \ttbar\ and $WW$ backgrounds is well achieved by \mttwo\ which is bounded from above by the mass of $W$ boson. The \mttwo\ distribution is shown in the ATLAS \textit{2/3-lepton} search\cite{23L_conventional} in signal region requiring two leptons (left). Limits obtained by the ATLAS \textit{2/3-lepton} and the \textit{RJigsaw}\cite{ATLAS_rjigsaw} analyses are compared for a simplified model with $\tilde{\chi}_{2}^{0}\tilde{\chi}_{1}^{\pm} \rightarrow  W (\rightarrow l\nu /qq') \tilde{\chi}_{1}^{0} Z (\rightarrow ll)\tilde{\chi}_{1}^{0}$}
\end{figure}

As an alternative to this search, the \textit{RJigsaw} analysis\cite{ATLAS_rjigsaw} studied the same models, using \textit{RJigsaw} variables. These variables assume a specific \textit{decay tree} (i.e. a specific decay pattern) of charginos and neutralinos. Different scale variables are constructed in every rest-frame of particles appearing in this decay tree by sorting visible and invisible particles:

\begin{equation}
 H_{n,m}^{\mathrm{F}} = \sum_{i=1}^{n} |\mathbf{p}_{\mathrm{vis},i}^{\mathrm{F}}| + \sum_{j=1}^{m} |\mathbf{p}_{\mathrm{inv},j}^{\mathrm{F}}|
\end{equation}

Examples for a decay tree and one of the variables are shown in Fig.~\ref{RJigsaw_variables}: $H_{n,1}^{\mathrm{PP}}$ is a scale variable in the rest frame of both initial variables (\ninotwo\ and \chinoone\ in this case) and acts in principle similar to \meff. 

\begin{figure}
 \centerline{
  \includegraphics[align=c,width=0.35\textwidth]{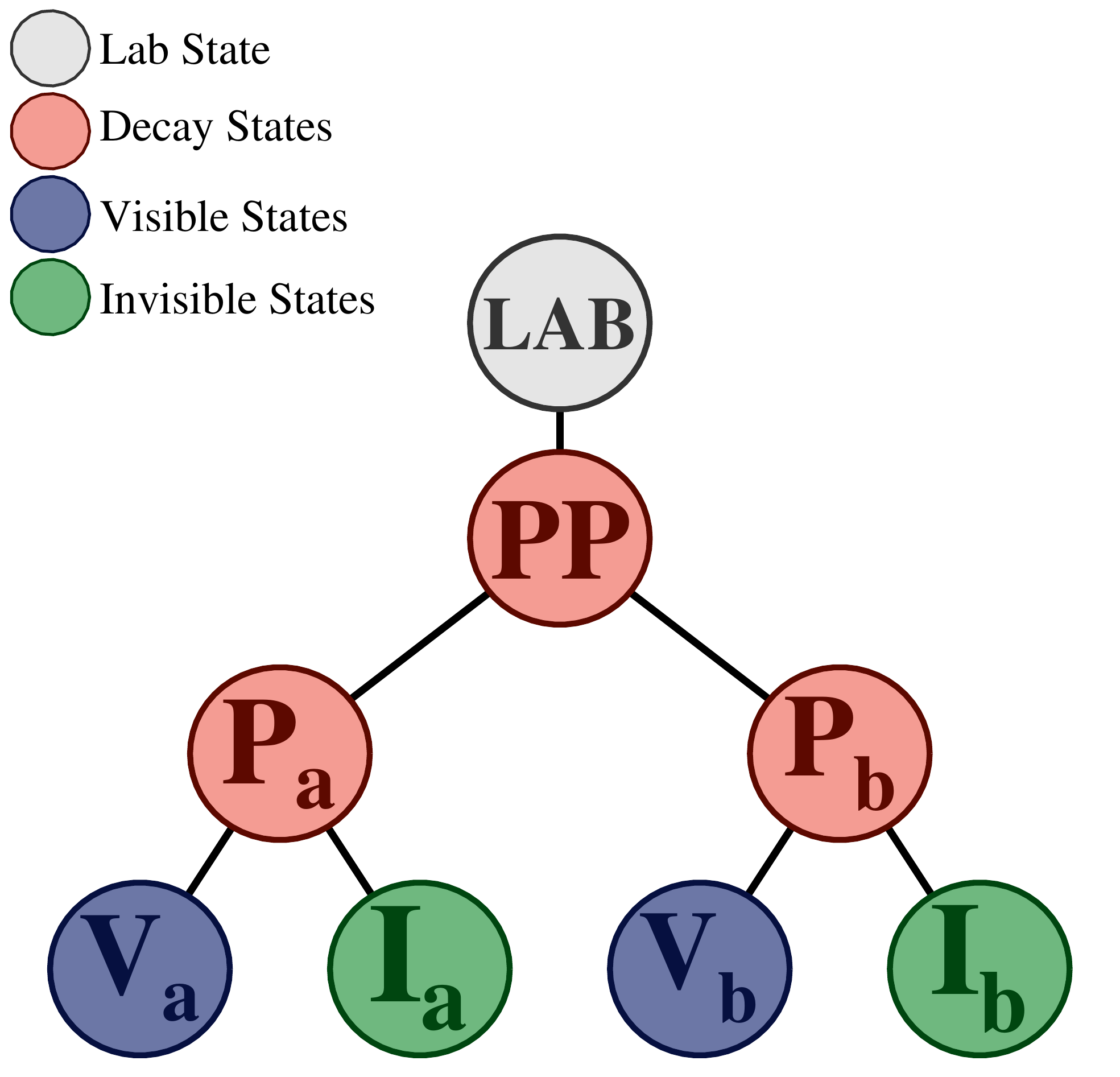}
  \includegraphics[align=c,width=0.48\textwidth]{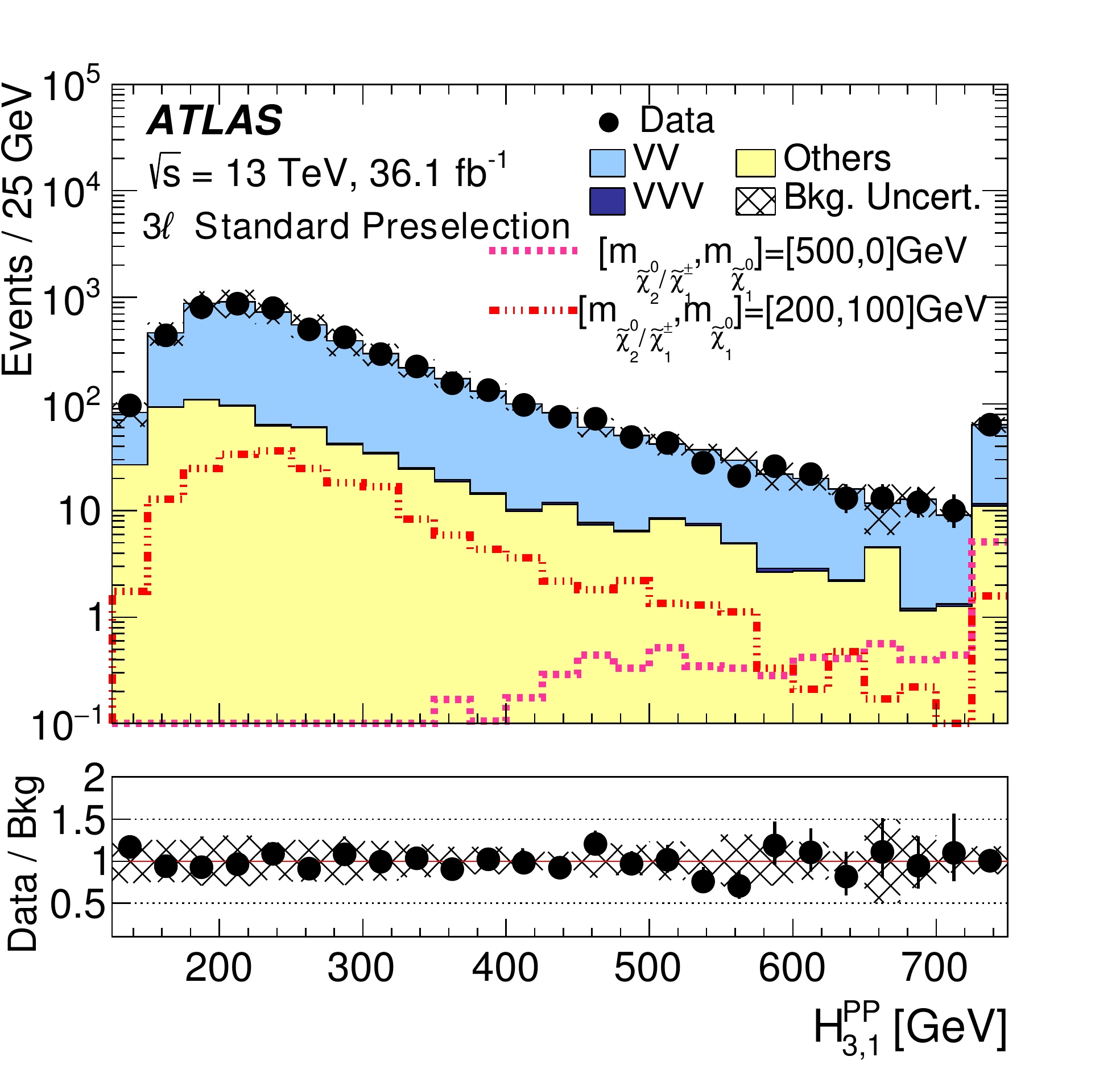}  
 }
 \vspace*{8pt}
\caption{\label{RJigsaw_variables}A standard decay tree is shown for the ATLAS \textit{RJigsaw} analysis\cite{ATLAS_rjigsaw} (left). Starting with the rest frame of the initially produced sparticles, particles are sorted into visible and invisible particles in the final state. RJigsaw variables are constructed based on each rest frame at each level of the decay tree. An example for a RJigsaw scale variable assuming a decay tree with three leptons in the final state is shown on the right.}
\end{figure}

The \textit{RJigsaw} analysis defines four different sets of signal regions with either two or three leptons and addressing signal scenarios with small $\Delta m$ or larger $\Delta m$ as shown in Fig.\ref{RJigsaw_results}. Few excesses in the order of $3 \sigma$ are observed in signal regions targeting small $\Delta m$ or signatures with ISR jets. The exclusion limit at 95 \% CL is shown in Fig.~\ref{RJigsaw_results} where it is compared to the conventional \textit{2/3-lepton} search. The two analyses select entirely different events in their signal regions, and are thus complementary, despite still targeting the same signal simplified models. A recent conference contribution\cite{RJigsaw_CONF} presented a re-implementation of the \textit{RJigsaw} analysis using conventional variables. While being able to reproduce the \textit{RJigsaw} analysis and results using the partial dataset of $36.1$ \ifb, the excess could not be confirmed in the full dataset of $139$ \ifb.

\begin{figure}
 \centerline{
  \includegraphics[width=0.49\textwidth]{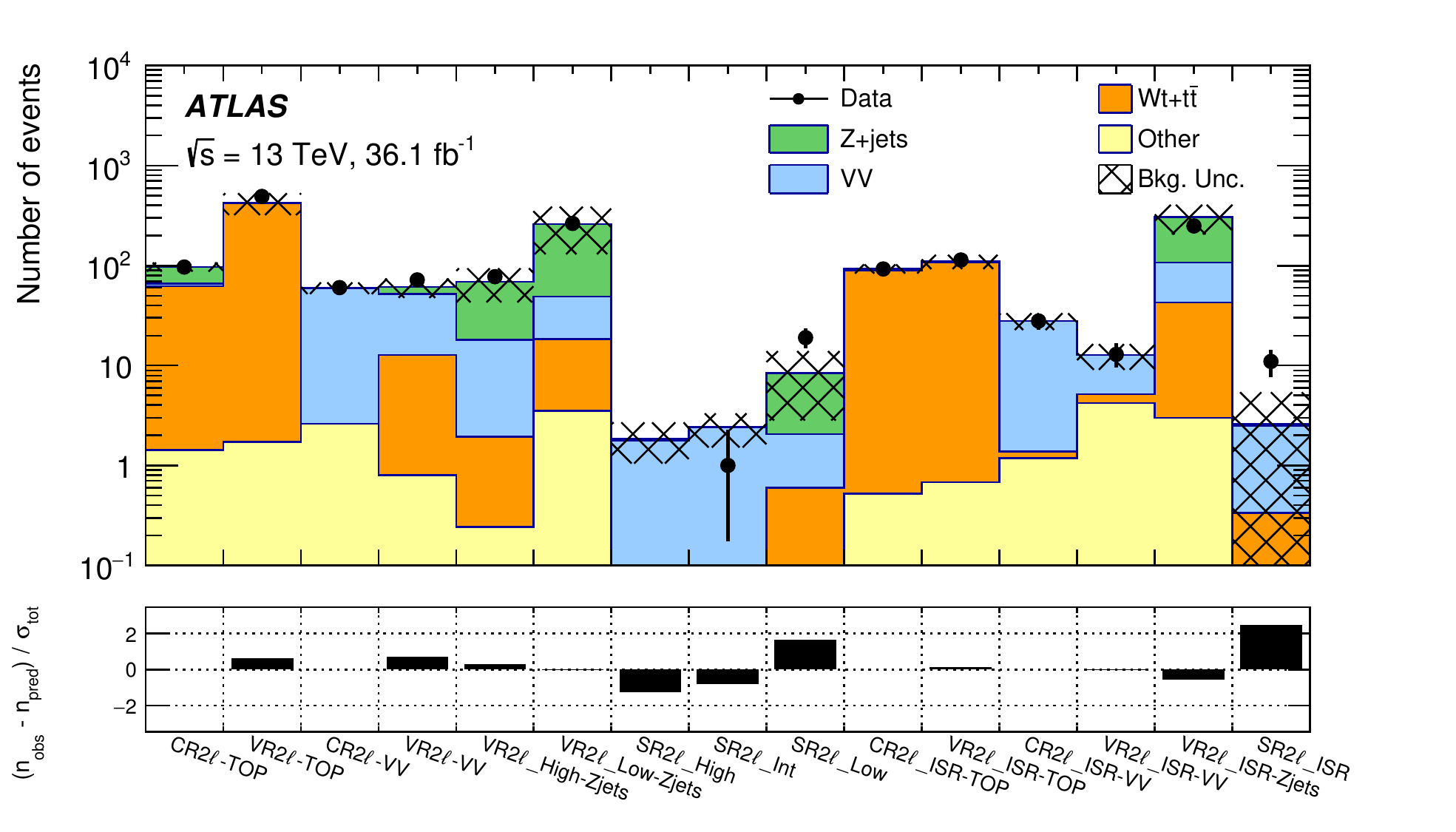}
  \includegraphics[width=0.49\textwidth]{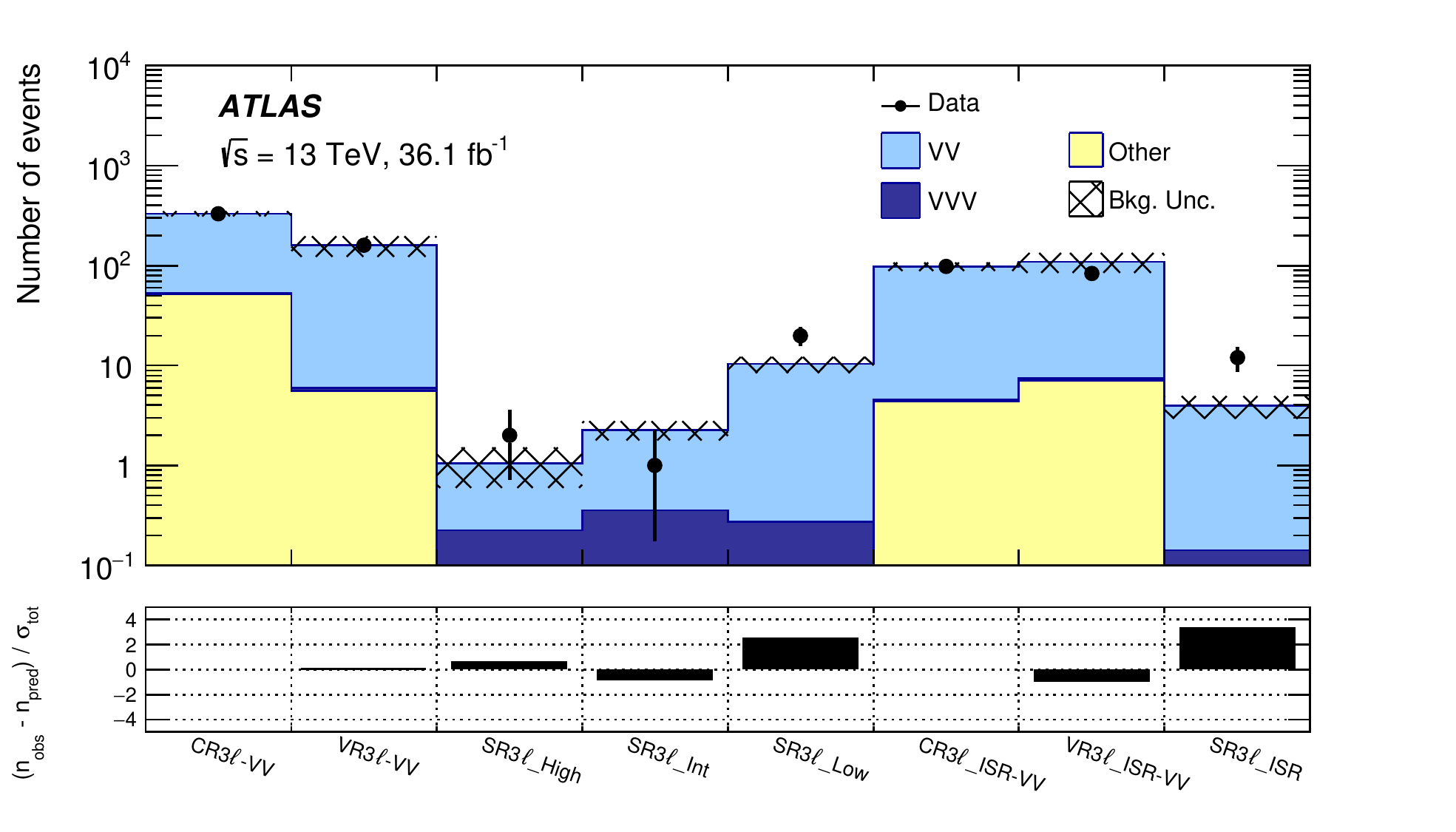}  
 }
 \vspace*{8pt}
\caption{\label{RJigsaw_results}Comparison of the background estimates with the observed data in the ATLAS \textit{RJigsaw}\cite{ATLAS_rjigsaw} signal regions. In four signal regions a modest excess was observed.}
\end{figure}

Decays of \ninotwo\ to a Higgs boson and \ninoone\ are preferred for many choices of parameters in the MSSM if the mass difference is larger than the mass of the Higgs boson and assuming that higgsinos are heavier than winos. Searches for such decays were performed assuming a simplified model with pair production of mass-degenerate \ninotwo \chinoone\ and $\tilde{\chi}_{2}^{0}\tilde{\chi}_{1}^{\pm} \rightarrow  h \tilde{\chi}_{1}^{0} W \tilde{\chi}_{1}^{0}$ by both the ATLAS and CMS Collaborations\cite{ATLAS_Wh,CMS_WhWZ}.
The signature depends of the decay products of $W$ and Higgs bosons. Searches were performed in final states including two photons (from the Higgs decay) and one lepton, in fully hadronic final states including two b-tagged jets, in a final state with a lepton and two b-tagged jets and in multi-lepton signatures. These different analyses provide a good complementarity as illustrated in Fig.~\ref{EWK_summary}. The \textit{diphoton + lepton} analysis\cite{ATLAS_1Lyy} profits from a very pure final state and can thus estimate the backgrounds via a side-band fit to falling background distributions. With this, the analysis is able to cover low \ninotwo\ and \chinoone\ masses. The \textit{1-lepton + $b\bar{b}$} analysis\cite{ATLAS_1Lbb} covers larger \chinoone\ and \ninotwo\ masses. Due to a sophisticated multi-dimensional fit of exclusive signal region bins, this preliminary analysis is currently able to reach the best limits in this model. Masses of degenerate \chinoone\ and \ninotwo\ of up to 740 \GeV\ can be excluded for small \ninoone\ masses. A fully-hadronic analysis\cite{ATLAS_Wh} requiring just two b-tagged jets was performed in context of this model for the first time. It provides better limits to high \chinoone\ and \ninotwo\ masses if considering the same dataset. 

\begin{figure}
 \centerline{
  \includegraphics[width=0.56\textwidth]{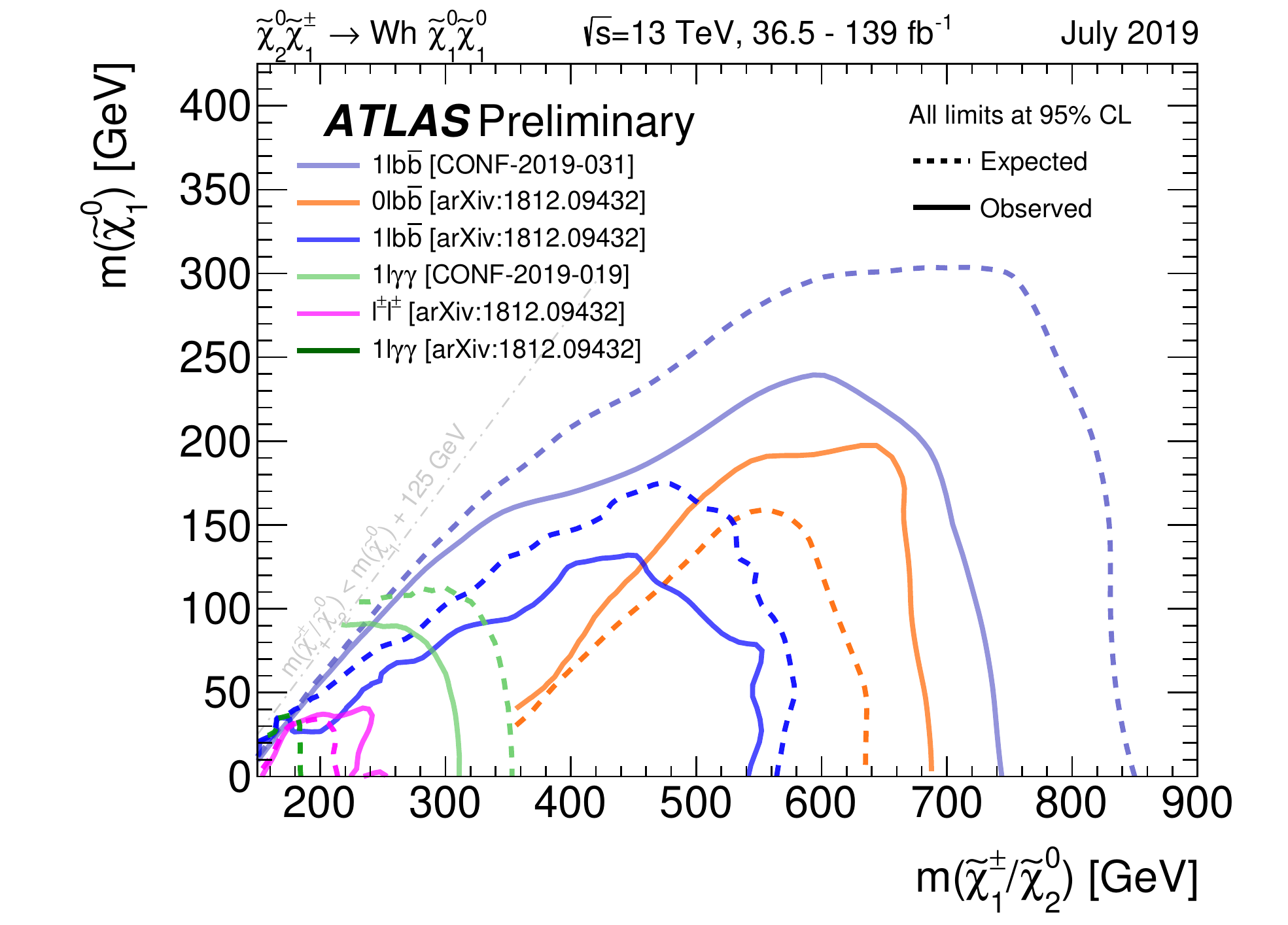}
  \includegraphics[width=0.42\textwidth]{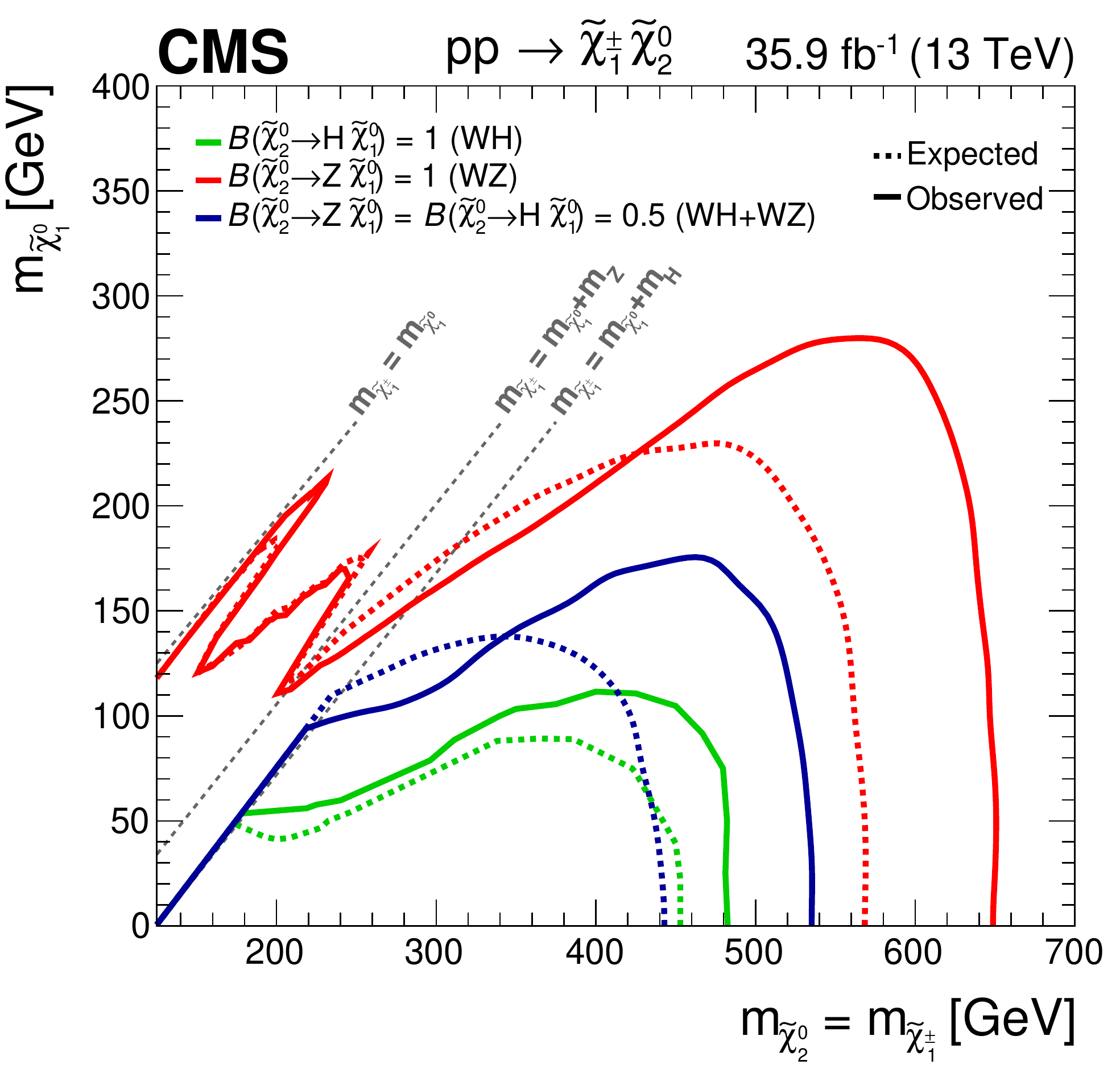} 
 }
 \vspace*{8pt}
\caption{\label{EWK_summary}Summary of exclusion limits\cite{stop_summary_plot} at 95 \% CL by the ATLAS Collaboration for the simplified model $\tilde{\chi}_{2}^{0}\tilde{\chi}_{1}^{\pm} \rightarrow  h \tilde{\chi}_{1}^{0} W \tilde{\chi}_{1}^{0}$ (left). Summary of exclusion limits\cite{CMS_WhWZ} by the CMS Collaboration for the simplified models $\tilde{\chi}_{2}^{0}\tilde{\chi}_{1}^{\pm} \rightarrow  h/Z \tilde{\chi}_{1}^{0} W \tilde{\chi}_{1}^{0}$ (right).}
\end{figure}

The CMS Collaboration presented a combination\cite{CMS_WhWZ} of all their searches for decays of a \ninotwo\chinoone-pair into $W$, $Z$ and Higgs bosons with a partial Run-2 dataset of 35.9 \ifb. The exclusion limits are reported in Fig.~\ref{EWK_summary}.

\subsubsection{Searches for sleptons}

Staus can play a role in the co-annihilation with neutralinos. Light staus may lead to the correct DM relic density. Searches for staus are challenging due to yet smaller cross sections than for neutralinos and charginos. The full dataset of Run-2 allows to reach sensitivity to staus for the first time. A preliminary analysis\cite{ATLAS_stau} by the ATLAS Collaboration excludes stau masses between 120 and 390 \GeV\ assuming pair-production of $\tilde{\tau}\tilde{\tau}$ with $\tilde{\tau} \rightarrow \tau \tilde{\chi}_{1}^{0}$.

\subsubsection{Summary of searches for electroweakinos}

Both the ATLAS and the CMS Collaborations present a comprehensive search program for charginos, neutralinos and sleptons.
All of these searches are based on simplified models and the limits obtained are not directly applicable to the MSSM.

Some of the searches for electroweakinos performed by ATLAS, CMS (using the 2015+2016 data) and also LEP were re-interpreted together with constraints on invisible decays of $Z$- and Higgs bosons by the GAMBIT Collaboration in a likelihood combination\cite{gambit_paper}. The authors conclude that no mass range for charginos and neutralinos can robustly be excluded in the MSSM by the ATLAS and CMS searches considered, since none of the searches is sensitive to more complex decay patterns of charginos and neutralinos than assumed in the simplified models.

\section{Comparison of DM searches in the SUSY context}
\label{SUSYDMDD}

Using the dataset of Run-1, searches for electroweakinos considering signatures with two to four charged leptons were interpreted in a reduced version of the pMSSM with only the parameters relevant to electroweakinos appearing\cite{Aaboud:2016wna}. The findings were compared to results of DD experiments, illustrating the complementarity in particular for regions with $m(\ninoone) \lesssim 65 \GeV$ and spin-independent interaction cross-sections\cite{Aaboud:2016wna}. Most of the DM models in the $Z$- or $h$-funnel regions with $m(\ninoone) \approx 45 \GeV$ or $m(\ninoone) \approx 65 \GeV$, respectively, were excluded, while only limited constraints could be made on the co-annihilation case with $m(\ninoone) \approx m(\chinoone)$ or the A-funnel region with $0.2 \lesssim m(\ninoone) \lesssim 2 \TeV$. The Run-2 search program at ATLAS and CMS however closed many of the short-comings of the Run-1 search program, so a repetition of the studies using the latest results will be of interest. 

\section{Searches for BSM mediators decaying to invisible states}
\label{monoX}

In generic searches for BSM mediators decaying to invisible states, the production of DM particles is tagged by the emission of a SM particle from the initial state, like a gluon (resulting in mono-jet signatures), photon, or $V$ boson, or by the emission of SM particles like e.g. a Higgs boson from the mediator particle. This emission is recoiling against the created DM particles, causing them to result in \met. 

These searches are usually interpreted in BSM mediator simplified dark matter models and in some cases also in SUSY models, as mentioned in Sec.~\ref{SUSYsearches}. In the following a few examples of these searches are highlighted.

\subsection{jet/$V$+\met}

Searches for a jet or a vector boson + \met\ provide sensitivity to a variety of different DM models, e.g. to the (axial/pseudo)-vector/scalar mediator models, but also to SUSY models, and thus serve as important benchmark analyses. In particular the jet +\met\ analysis profits from frequently occurring initial-state radiation of gluons at hadron colliders. The \textit{jet/$V$+\met} search\cite{CMS_monojet} by the CMS Collaboration is one recent example (using 35.9 \ifb) of these searches\cite{ATLAS_monojet,CMS_monojet2,ATLAS_monojet2}. This analysis selects events with a relatively large \met\ and $H_{\mathrm{T}}^{\mathrm{miss}}$ constructed from the vector \pt\ of the jets present in the event. At least a high-energetic jet is required in the central detector region and the presence of leptons and photons vetoed. By further rejecting events with b-tagged jets, a suppression of the \ttbar\ background is achieved. Other important backgrounds are $Z \rightarrow \nu\nu$ and $W \rightarrow l\nu$. In case of the \textit{$V$+\met} signal regions, further criteria on jets with a large cone size are imposed to reconstruct hadronically decaying $W$ bosons.  
No significant excess is observed. Exclusion limits may reach up to 1.8 \TeV\ on the mediator mass in (axial-)vector simplified Dark Matter models assuming $g_{q} = 0.25$, $g_{l} = 0$ and $g_{\chi} = 1$. Limits are compared to DD experiments as presented in Fig.~\ref{monoX_limits}. This comparison is only valid under assumption of this specific model. For DM masses below 5 \GeV\ the exclusion limits turn out to be significantly stronger than the DD limits. The \textit{jet/$V$+\met} search thus presents a complementarity reach with respect to DD experiments.

\begin{figure}
 \centerline{
  \includegraphics[width=\textwidth]{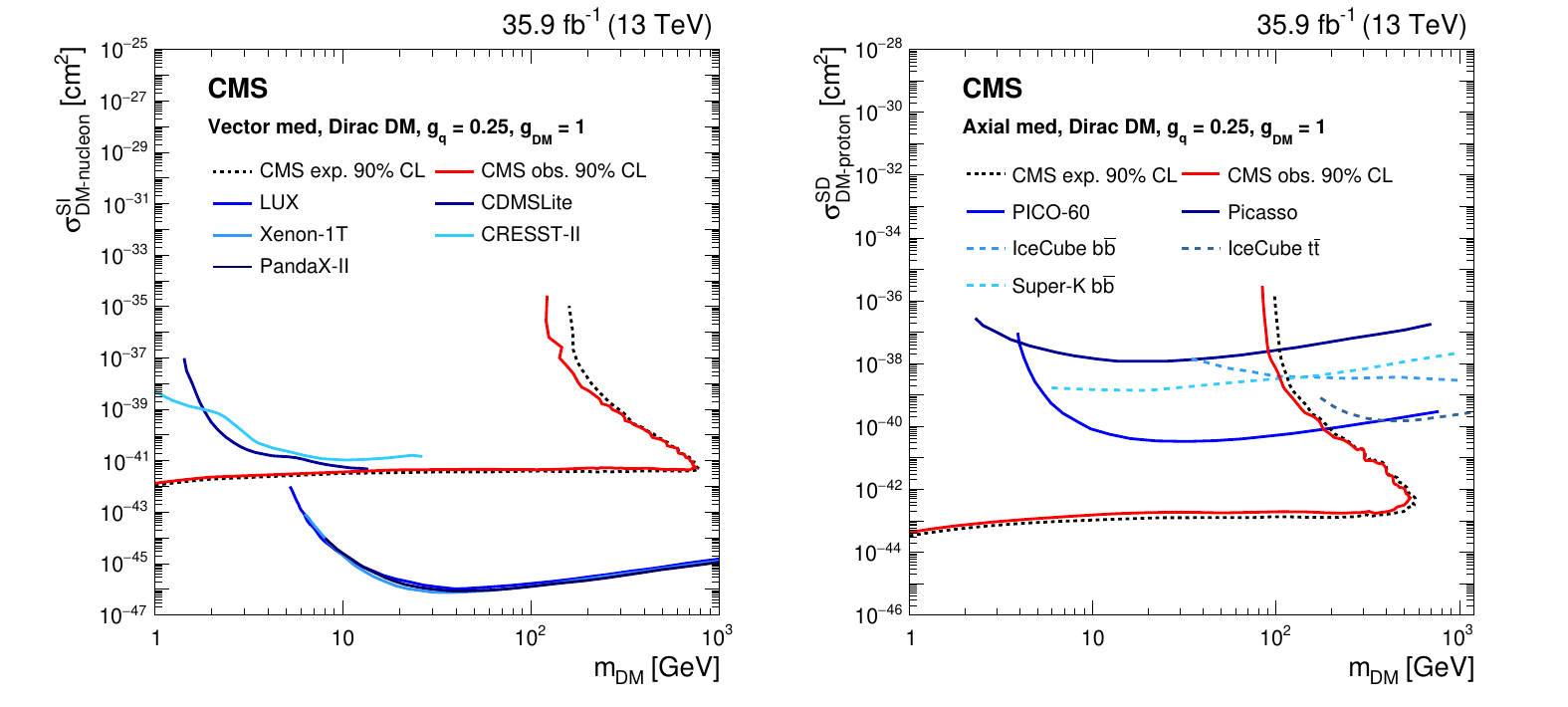}
 }\vspace*{8pt}
\caption{\label{monoX_limits}Limits in the $m_{\mathrm{DM}}$ versus $\sigma_{\mathrm{SI/SD}}$ plane for vector (left) or axial-vector (right) simplified models obtained by the \textit{jet/$V$+\met} search\cite{CMS_monojet} at 90 \% CL in comparison to limits by direct-detection experiments. SI refers to spin-independent and SD to spin-dependent.}
\end{figure}

Using this signature, limits are also set on production of SUSY particles like gluinos, stops and sbottoms\cite{ATLAS_monojet}, as discussed in Sec.~\ref{SUSYsearches}.

\subsection{$\gamma$+\met}

Similarly to the \textit{jet+\met} searches, the \textit{$\gamma$+\met} signature offers the possibility to consider SUSY models or DM models without assumptions on the underlying theory.

In the \textit{$\gamma$+\met} analysis\cite{CMS_photon} by the CMS Collaboration (using 35.9 \ifb) a high-energetic isolated photon along with large \met\ is required. Additional criteria are applied to suppress backgrounds resulting in leptons or originating from beam-halo or mis-measurements. In the (axial-)vector simplified model, limits on the mediator mass of 950 \GeV\ are excluded for small DM masses, assuming $g_{q} = 0.25$, $g_{l} = 0$ and $g_{\chi} = 1$. A similar search exists from ATLAS\cite{ATLAS_photon}, setting limits up to 1.2 \TeV\ on the mediator mass for small DM masses in the same models.

\subsection{$h$+\met}

An initial-state radiation of a Higgs boson is Yukawa-suppressed due to the low fraction of heavy flavor quarks in the proton and the heavy mass of the Higgs boson. Searching for a $h$+\met\ signature thus probes the emission of a Higgs boson from the mediator particle and thus directly the connection of the Higgs sector with a dark sector\cite{doglioni,ATLAS_DMsummary}. Such Higgs emission is possible in different models, e.g. in a type-II 2HDM model\cite{2HDM_1,2HDM_2,2HDM_3} with additional U(1) gauge symmetry. In this case, the $h$ is created in the decay of the mediator via $Z'_{V} \rightarrow h A (\rightarrow \chi\bar{\chi})$ as shown in Fig.~\ref{monoX_diagrams}, middle. A particular sensitive search with good background suppression is achieved by considering decays $h \rightarrow b\bar{b}$, which gives rise a distinctive final state with $b\bar{b}$ and \met. Various searches have addressed this final state\cite{ATLAS_monoh_paper,ATLAS_monoh_CONF,CMS_monoh,CMS_monoh3}, while other searches -- without reaching similar sensitivity -- have also probed other Higgs decays in this context\cite{CMS_monoh2,CMS_monoh3,ATLAS_monohyy}. The recent ATLAS \textit{$h$+\met} search\cite{ATLAS_monoh_CONF}, using 80 \ifb, defines different exclusive signal regions depending on if the two b-tagged jets are as boosted as to merge into a jet with large cone radius (merged region with large cone size jet) or not (resolved region with two small cone size jets). Events with leptons are rejected to achieve a good background suppression of the dominant \ttbar\ and $W$+jets backgrounds. Further requirements on the multiplicity of b-tagged jets are made, and different exclusive search regions in the mass of the large cone size jet or the invariant mass of the two small cone size jets defined in order to reconstruct the Higgs mass. This analysis uses two novel methods to improve its sensitivity further. By constructing an object-based \met\ significance\cite{ATLAS_metsig}, a measure is provided on how likely the \met\ originates from invisible particles in contrast to mis-measurements from resolution effects. This variable is constructed by explicitly considering the longitudinal variance $\sigma_{L}$ and the correlation $\rho_{LT}$ between longitudinal and transverse components of all objects contributing to \met:

\begin{equation}
 S = \sqrt{\frac{|\mathbf{E}_{\mathrm{T}}^{\mathrm{miss}}|^{2}}{\sigma_{L}^{2}(1 - \rho_{LT}^{2})}}
\end{equation}

\noindent The performance of this variable was found to be superior\cite{ATLAS_monoh_CONF} to more traditional variables like \met/$\sqrt{\sum_{\mathrm{jets}} p_{\mathrm{T}}}$ (event-based \met\ significance) as shown in Fig.~\ref{monoh_plots}.

\begin{figure}
 \centerline{
  \includegraphics[width=0.4\textwidth]{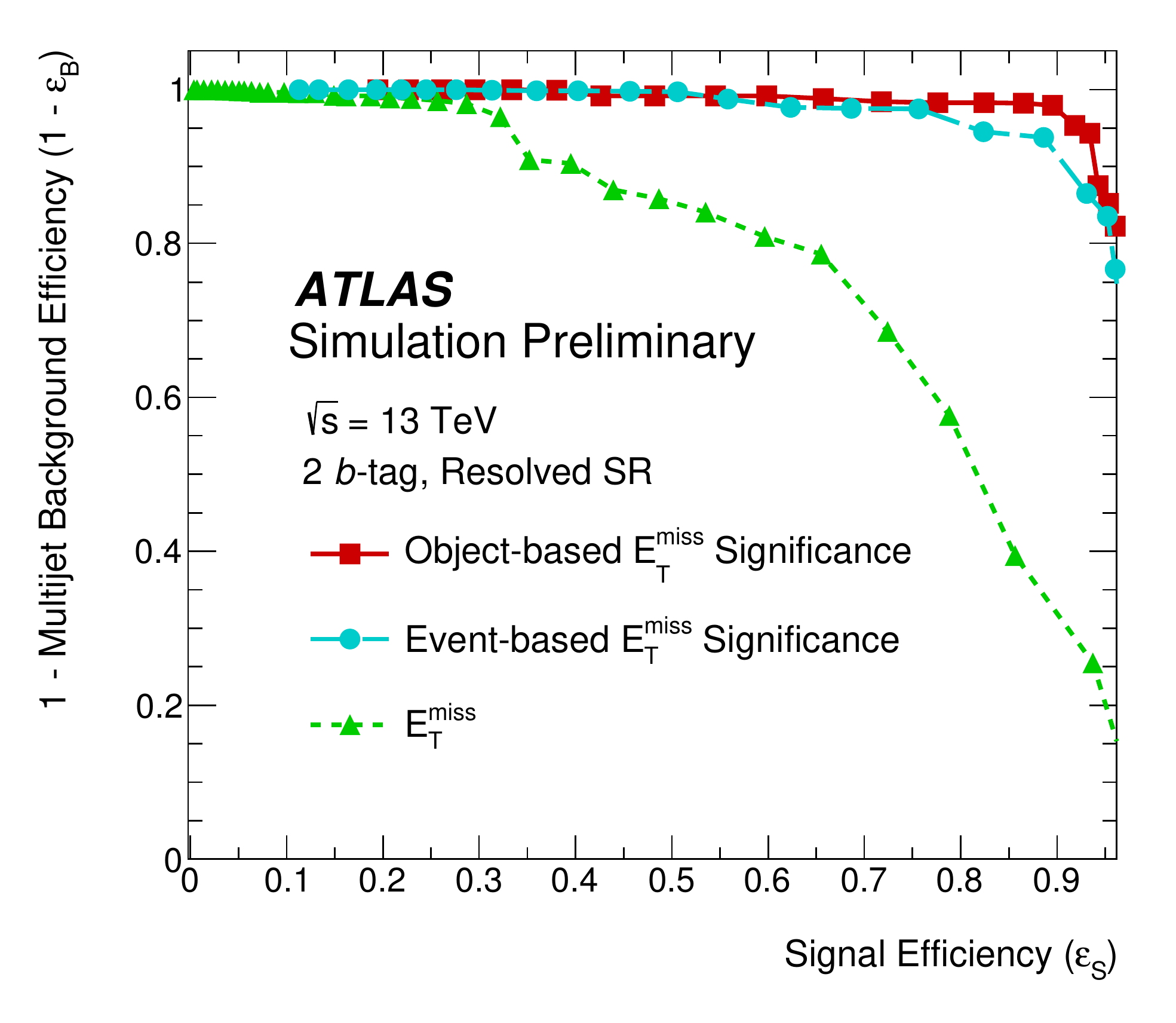}
  \includegraphics[width=0.49\textwidth]{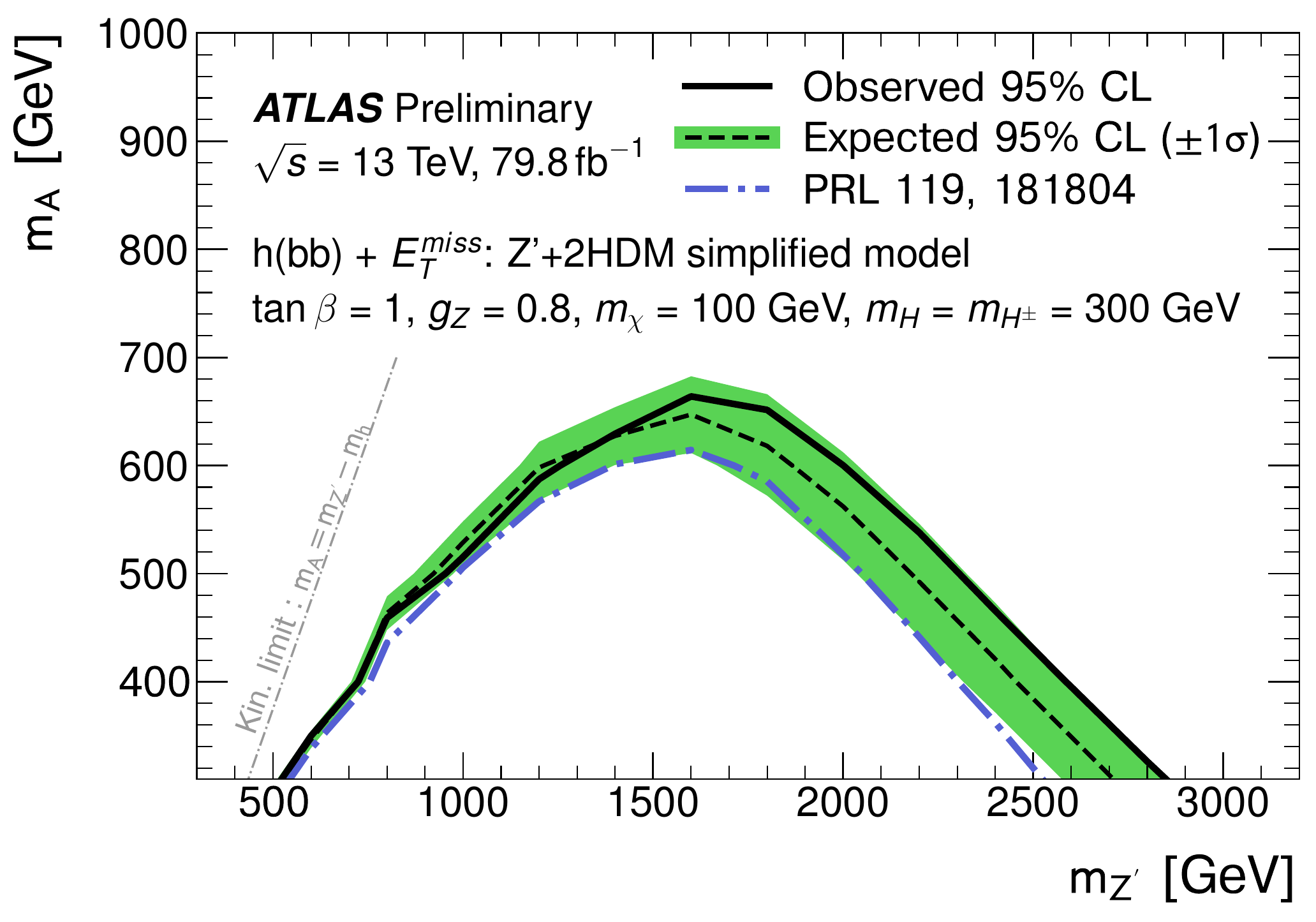}
 }\vspace*{8pt}
\caption{\label{monoh_plots} 
The ATLAS \textit{$h$+\met} analysis\cite{ATLAS_monoh_CONF} improves its sensitivity by using the object-based \met\ significance instead of the event-based significance (left). Definitions see text. Limits at 95 \% CL are given in a type-II 2HDM model with additional U(1) gauge symmetry (right).}
\end{figure}

In the merged region a special jet definition using a variable cone size gives superior performance, as this allows to improve the identification of $b$-quarks from boosted Higgs decays. The analysis sets limits on the mediator mass $m_{Z'_{V}}$ up to 2.8 \TeV\ depending on the specific signal model as shown in Fig.~\ref{monoh_plots}.

\section{Implications of precision measurements of the Standard Model for BSM physics and Dark Matter}
\label{SM}

The Run-2 dataset allows to perform a comprehensive program of precision measurements of SM processes.
Measurements of SM processes are also a test of potential BSM contributions. Properties of the top quark are particular sensitive to BSM effects, as the top quark plays a special role in many BSM theories due to its high mass and large coupling to the Higgs boson and has thus a connection with the problem of naturalness, as e.g. summarized in Ref.~\refcite{topreview}. For example, measurements of the \ttbar\ spin correlation\cite{ttbar_spin_ATLAS,ttbar_spin_CMS} can be interpreted as constraints on $\tilde{t}_{1} \rightarrow X \ninoone$, excluding stop masses between 170 and 230 \GeV\ for different \ninoone-masses\cite{ttbar_spin_ATLAS}.

Precision measurements of quantities related to the Higgs boson allow to test various SM extensions. For instance, measurements of the various Higgs production and decay modes allow to measure and constrain the couplings of the Higgs to (SM) particles\cite{ATLAS_couplings,CMS_couplings}. Deviations could hint at the presence of BSM physics. As indicated in Sec.~\ref{DMmodels}, in Higgs-portal DM models the Higgs boson might couple directly to DM particles, resulting in invisible decays of the Higgs to $\chi\bar{\chi}$. As invisible Higgs decays need to be tagged by the presence of some other SM particles, searches\cite{ATLAS_Hinv,CMS_Hinv} consider Higgs production channels with either additional jets (VBF topology), or $V$ bosons which may decay either into leptons or quarks. A combination of the different channels lead to a limit\cite{CMS_Hinv} at 95 \% CL of $B_{H \rightarrow \mathrm{invisible}} < 0.19$.
These constraints are interpreted as limits on the spin-independent DM-nucleon scattering cross-section as shown in Fig.~\ref{limits_hinv}.

\begin{figure}
 \centerline{
  \includegraphics[width=0.56\textwidth]{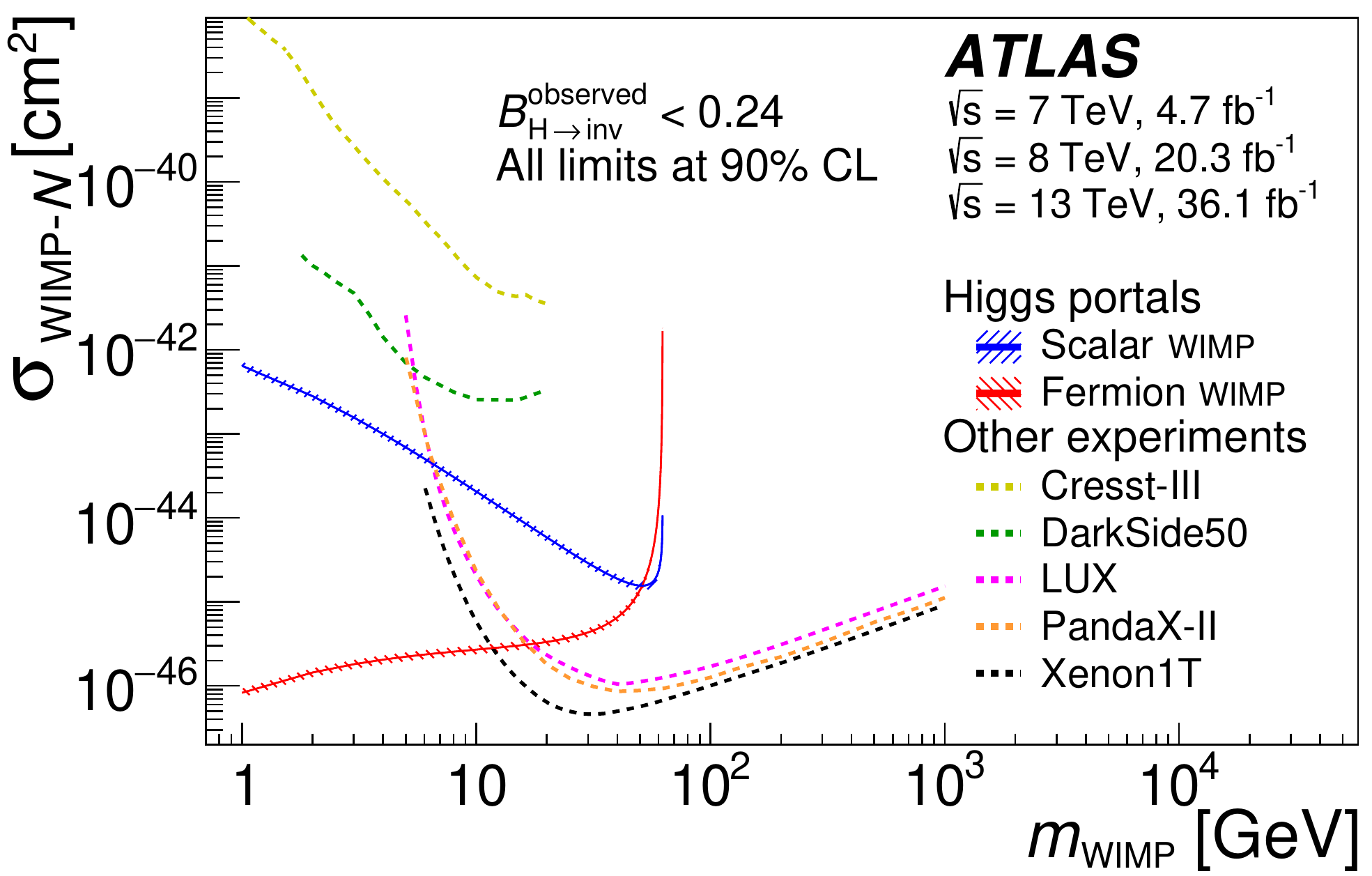}
  \includegraphics[width=0.42\textwidth]{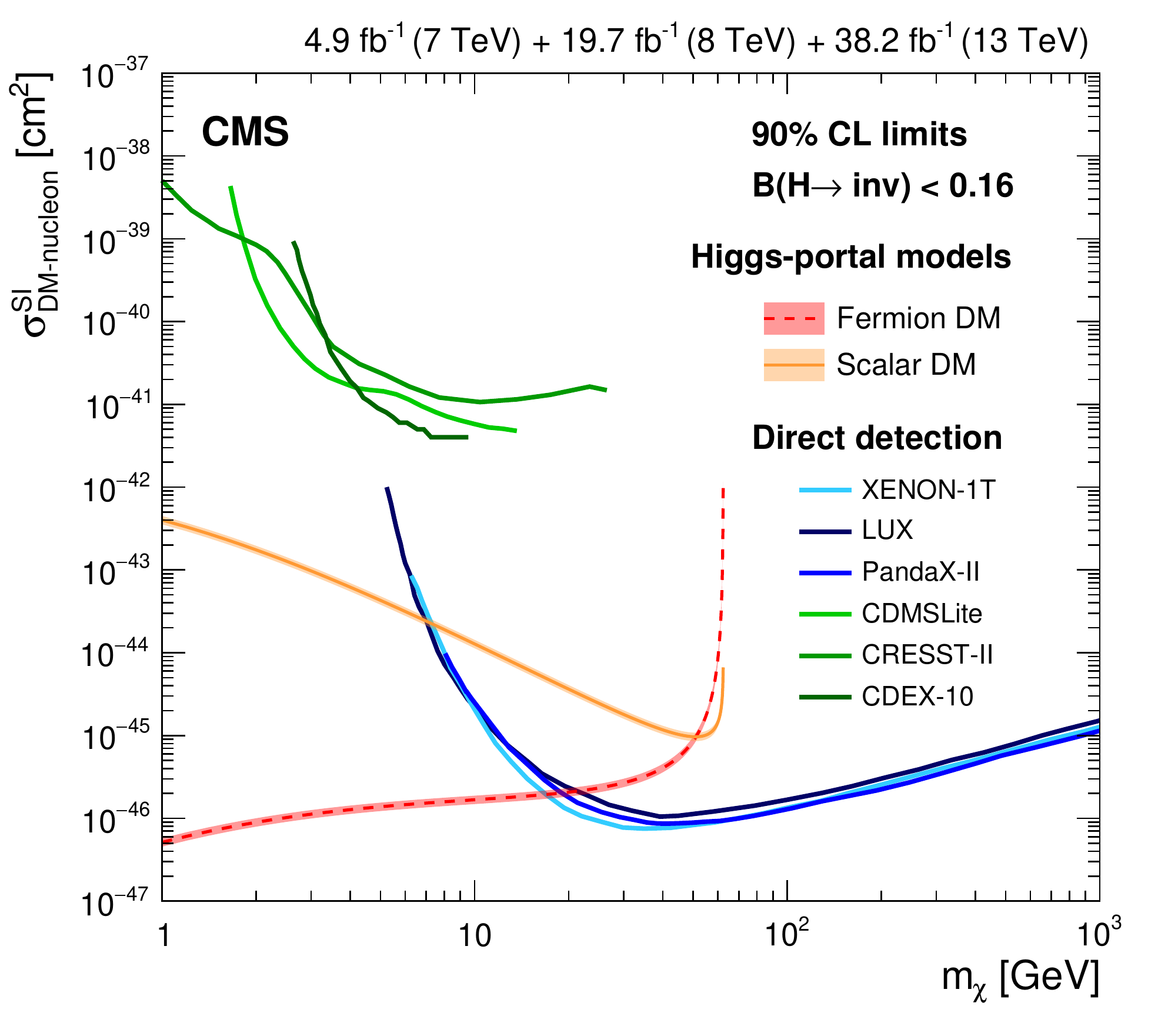}
 }\vspace*{8pt}
\caption{\label{limits_hinv}Searches for invisible Higgs decays can be interpreted as limits on the spin-independent DM-nucleon scattering cross-section. These limits provide complementary information to other experiments searching for DM\cite{ATLAS_Hinv,CMS_Hinv}.}
\end{figure}

\section{Comparison of DM searches in mediator simplified models}
\label{DMsummary}

Considering different searches for BSM mediators decaying to invisible states or SM particles, as well as interpretations of searches for stops/sbottoms and searches for invisible Higgs decays, constraints on a variety of mediator simplified dark matter models can be derived and compared to results by DD or ID experiments. Such a summary has been presented by ATLAS\cite{ATLAS_DMsummary} and by CMS\cite{CMSDMplots}. In comparing findings of different searches and experiments, not only specific models but also specific parameter values in these models have to be chosen. Comparisons are thus only valid under particular sets of assumptions. Benchmark models follow recommendations by the LHC DM WG\cite{Albert:2017onk,Boveia:2016mrp,Abe:2018bpo}. Fig.~\ref{DM_summary} presents two different interpretations\cite{ATLAS_DMsummary}. In case of a vector/axial-vector mediator simplified model, searches are compared assuming parameter values of $g_{q} = 0.25$, $g_{l} = 0$ and $g_{\chi} = 1$ (comparisons using other parameter values are made as well). Limits are presented as functions of $m_{Z'_{V}}$ and $m_{\chi}$. Searches for BSM mediators decaying to SM particles are typically sensitive to higher mediator masses $m_{Z'_{V}}$ in contrast to searches for BSM mediators decaying to invisible states reaching much lower masses. The strongest constraint in this model originates from a search for BSM mediators decaying to two jets (di-jet search). 

\begin{figure}
 \begin{center}
  \includegraphics[width=0.9\textwidth]{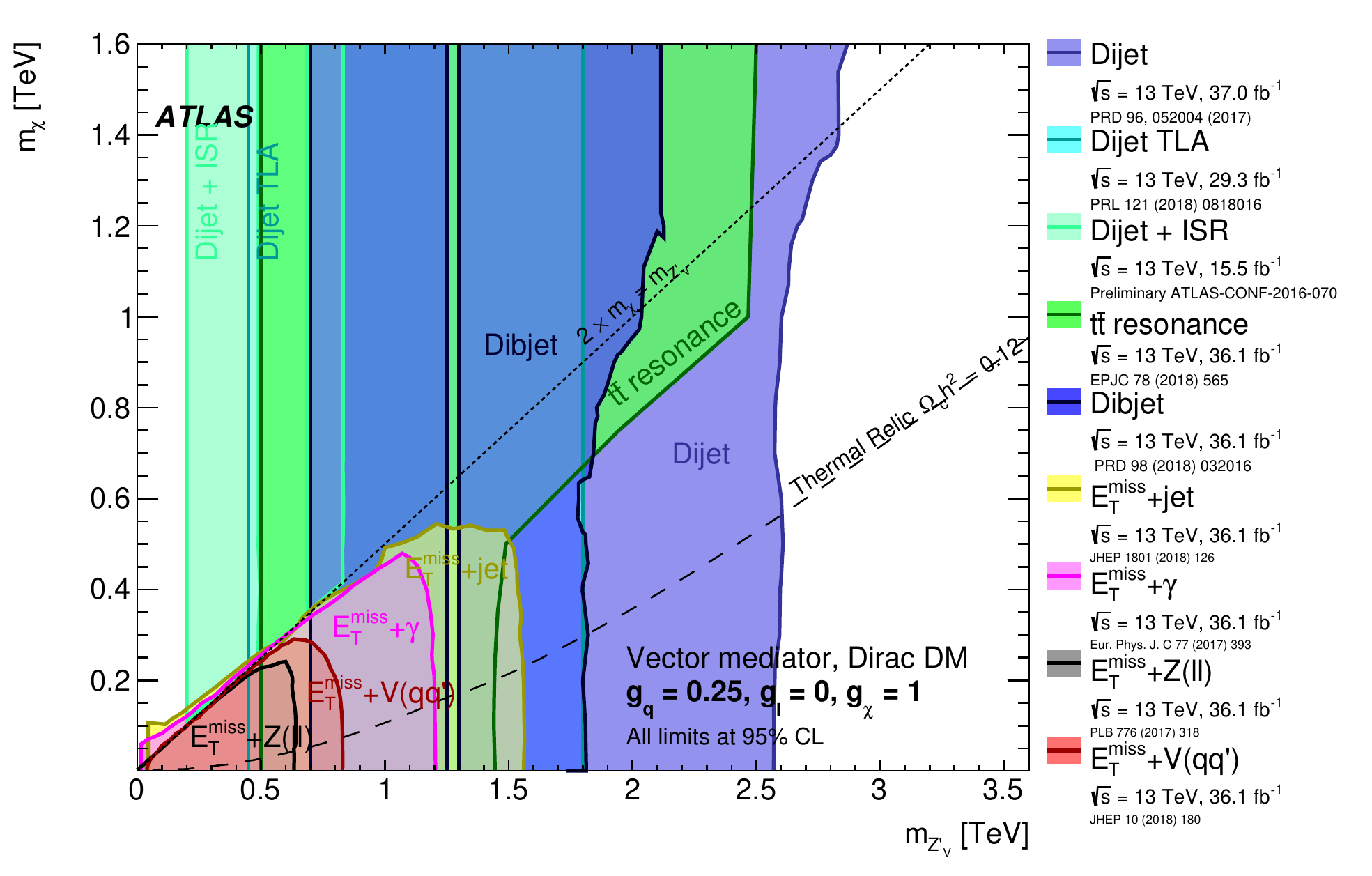}\\
  \includegraphics[width=0.9\textwidth]{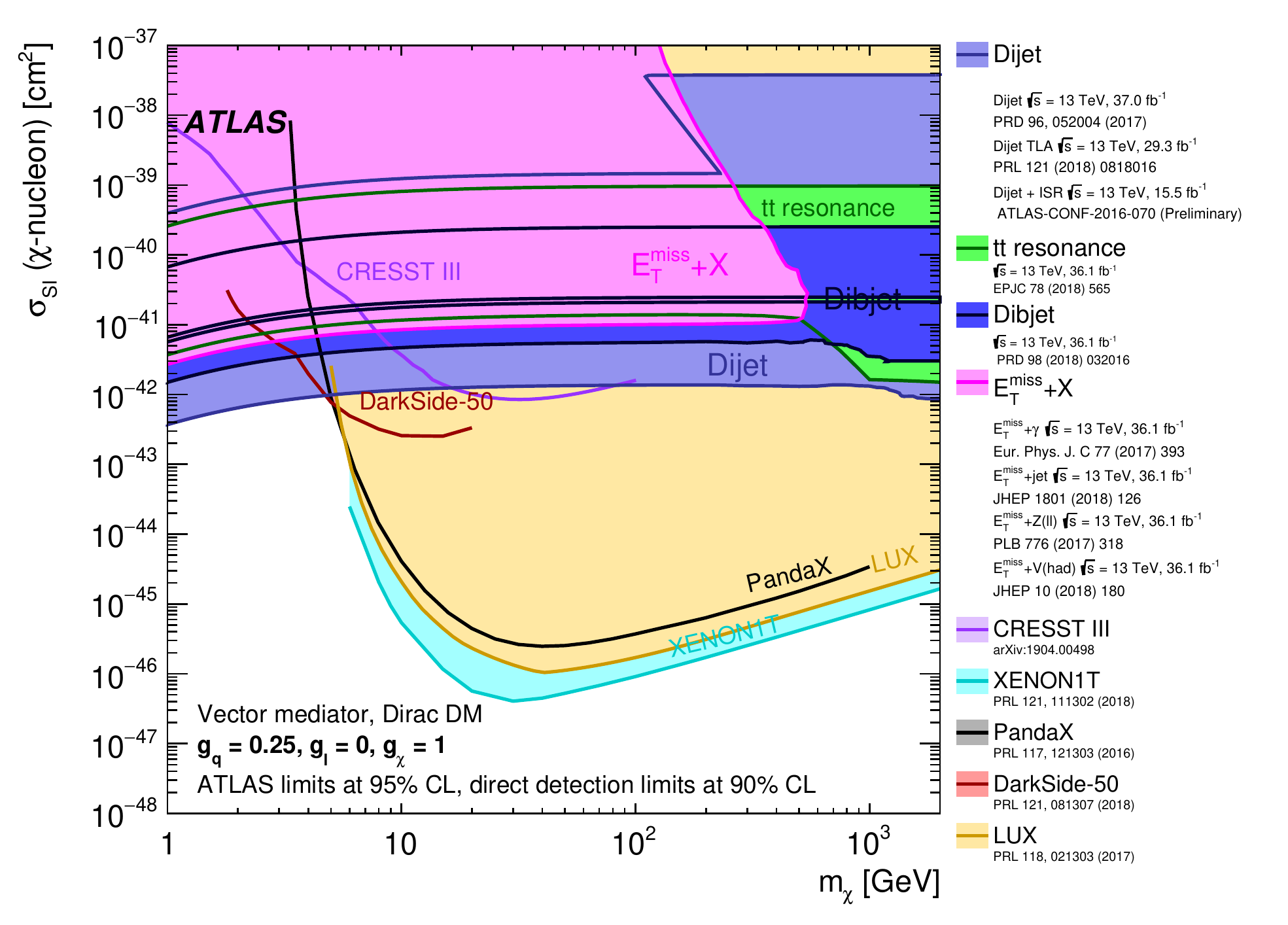}  
 \end{center}
\caption{\label{DM_summary}Summary of constraints\cite{ATLAS_DMsummary} in the vector-mediator simplified dark matter model with parameter settings $g_{q} = 0.25$, $g_{l} = 0$ and $g_{\chi} = 1 $ as function of $m_{Z'_{V}}$ and $m_{\chi}$ (top) and of $\sigma_{\mathrm{SI}}$ against $m_{\chi}$ (bottom).}
\end{figure}

Limits can also be displayed in the plane of $\sigma_{\mathrm{SI}}$ against $m_{\chi}$ assuming the vector mediator simplified model with the same parameter settings as before. Fig.~\ref{DM_summary} compares results by collider searches to findings of DD experiments. Collider experiments tend to be sensitive to lower $m_{\chi}$ than DD experiments. These plots demonstrate the complementarity of the different DM searches and experiments.

\section{Conclusion}

This review highlights a few searches for supersymmetric dark matter candidates and for DM appearing in mediator simplified models. In interpreting and comparing the results to DD experiments, it is important to consider the assumptions of the underlying model. None of these searches observed a significant excess.

Many innovative methods were employed to reach the best possible sensitivity. These methods include simultaneous likelihood fits of multiple exclusive search regions or methods of machine learning. Novel approaches to improve the reconstruction of particles, e.g. down to lower momentum, turned out to open up completely new territories for searches, as well as searches profited from improved \met\ reconstruction methods or jet substructure techniques. All these different innovations help to target challenging scenarios. Many of the searches presented do not use the full dataset of Run-2 yet -- the increased statistics will in particular help to obtain sensitivity to more difficult searches for charginos and neutralinos in the future. Searches for higgsino and wino LSPs have not reached the mass range suggested by relic density considerations of about 1.1 \TeV\ or 3 \TeV\ for pure higgsinos or winos, respectively, yet. Future colliders will be necessary to cover this mass range.

Dark matter particles may hide in even more difficult scenarios to address, like not promptly-decaying (long-lived) particles, which are not considered by most conventional searches. A comprehensive search program at colliders for all the different possibilities, complementing DD and ID experiments, will help in closing in on DM.

\section*{Acknowledgments}

The author would like to thank Monica d'Onofrio, Marie-H\'el\`ene Genest, Federico Meloni and Max Swiatlowski for providing very useful comments on the draft.

\end{document}